\documentclass[aps,twocolumn]{revtex4}%
\pdfoutput=1
\usepackage{amsfonts}
\usepackage{amsmath}
\usepackage{amssymb}
\usepackage{graphicx}%
\begin{document}
\title{Singularities in Speckled Speckle: Statistics }
\author{Isaac Freund}
\author{David A. Kessler}
\affiliation{Physics Department, Bar-Ilan University, Ramat-Gan IL52900, Israel}
\date{22 June, 2008}

\begin{abstract}
Random optical fields with two widely different correlation lengths generate
far field speckle spots that are themselves highly speckled. We call such
patterns \emph{speckled speckle}, and study their critical points
(singularities and stationary points) using analytical theory and computer
simulations. \ We find anomalous spatial arrangements of the critical points
and orders of magnitude anomalies in their relative number densities, and in
the densities of the associated zero crossings. \ 

\end{abstract}
\maketitle

\section{INTRODUCTION}

The canonical exemplars of highly structured light fields (\emph{complex
light}) are the random speckle patterns produced by scattering of coherent
light from surface or volume diffusers. \ Such patterns are not only of
special interest in their own right - in part because anything generic that
can happen in an optical field is likely to happen in a speckle pattern - but
also because of the myriad areas of practical application in which these
patterns appear [$1$].

\begin{figure}
[pb]
\begin{center}
\includegraphics[
natheight=3.442000in,
natwidth=1.959700in,
height=2.885in,
width=1.6544in
]%
{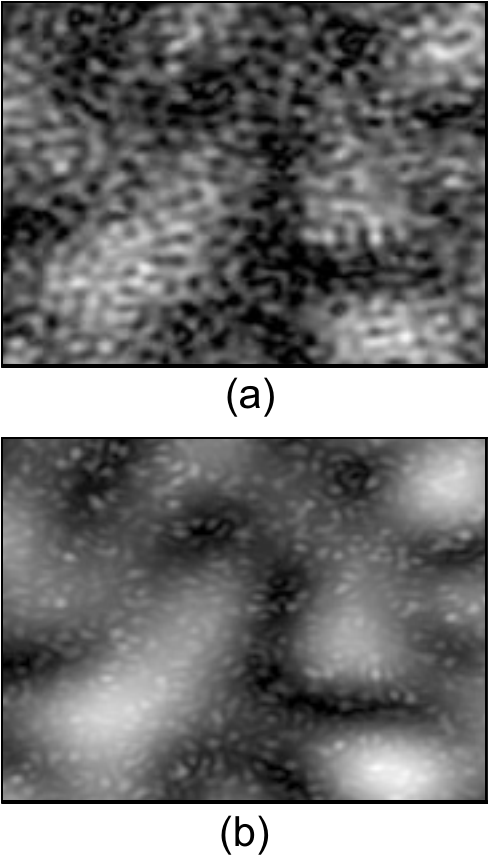}%
\caption{Speckled speckle. \ (a) Scalar field \ (single polarization
component) intensity. \ (b) Vector field (two polarization component)
intensity. The computer simulated source distributions that produce these
calculated speckle patterns consist of two concentric disks of light that
illuminate a random phase plate (diffuser). The diameters of these disks
differ by a factor of ten, and the total optical power in the smaller disk is
ten times the power in the larger disk. \ For the scalar field in (a) the
illuminating disks have the same linear polarizations, for the vector field in
(b) the polarizations are orthogonal. \ Individually, the small (large)
diameter disk produces large (small) speckle spots. Collectively, in (a)
interference between the speckle fields generated by the small and large disks
produces the fine granulation of the large speckle spots, whereas the pattern
in (b) is the sum of the speckle intensities generated by each disk because
orthogonal polarizations do not interfere. \ This absence of interference
notwithstanding, as discussed in Section III, the polarization singularity
structure of the vector field in (b) differs fundamentally from the
corresponding singularity structure produced by each disk acting alone.}%
\label{Fig1}%
\end{center}
\end{figure}

Prior theoretical and experimental studies of singularities in speckle
patterns have concentrated on fields with a single characteristic length scale
[$1$]. \ Here we study speckle patterns with two widely different length
scales. \ We call such patterns \textquotedblleft speckled
speckle\textquotedblright, because as shown in Fig. \ref{Fig1}, the major
speckle spots in the pattern are themselves highly speckled. \ We study the
basic statistical properties of the singularities of these fields, in
particular their relative number densities, and associated level crossings,
and find anomalous, often surprising results.

The plan of this paper, which extends a previous brief report [$2$], is as
follows. In Section II we describe the composite source distributions that
generate the nine different speckled speckle fields studied here, and
calculate the autocorrelation functions of these fields. \ In Section III we
discuss phase vortices and phase extrema in scalar (one polarization
component) paraxial fields, and C points and azimuthal extrema in vector (two
polarization component) fields. \ In both cases we find highly anomalous
ratios for the number densities of singularities and extrema - ratios that can
differ by many orders of magnitude from those in normal speckle. \ In this
section we also consider level crossings of the derivatives associated with
these phase critical points, again finding large, orders of magnitude,
anomalies for the relative number densities. \ In Section IV we discuss
extrema, umbilic points, and the associated level crossings, of the real and
imaginary parts of the field, finding, once again, anomalously large ratios
for these number densities. \ We briefly summarize our main findings in
concluding section V.

Throughout, we use computer simulations to provide images of the underlying
field structures. \ Based on these simulations, which use realistic source
parameters, we conclude that the many unusual properties of speckled speckle
described here theoretically could also be studied experimentally.

\section{SOURCES AND AUTOCORRELATION FUNCTIONS}

We assume the (standard two dimensional) geometry shown in Fig. \ref{Fig2},
further assume circular Gaussian statistics [$1$] for the two dimensional,
paraxial, complex optical field $E\left(  x,y\right)  $ and its derivatives,
and write $E\left(  x,y\right)  $ in terms of its real $\mathcal{R}\left(
x,y\right)  $ and imaginary $\mathcal{I}\left(  x,y\right)  $ parts as
\begin{equation}
E\left(  x,y\right)  =\mathcal{R}\left(  x,y\right)  +\text{i}\mathcal{I}%
\left(  x,y\right)  . \label{ERI}%
\end{equation}
Invoking stationarity and circular symmetry, the normalized autocorrelation
function of the field in the plane of observation (the $xy$-plane) is defined
by
\begin{equation}
W\left(  v\right)  =\left\langle E^{\ast}\left(  0\right)  E\left(  v\right)
\right\rangle /\left\langle \left\vert E\right\vert ^{2}\right\rangle ,
\label{W}%
\end{equation}
where $\left\langle ...\right\rangle $ represents an ensemble average, and
\begin{equation}
v=\left(  x^{2}+y^{2}\right)  ^{1/2}.
\end{equation}
With these assumptions all\emph{ }one and two point statistical properties can
be obtained from $W\left(  v\right)  $ and its derivatives.

We take $E$ to be in the far field of a circularly symmetric planar source
$S\left(  u\right)  $, where $u$ measures radial position in the source plane.
\ The VanCittert-Zenike theorem [$1$] relates $W\left(  v\right)  $ and
$S\left(  u\right)  $ by
\begin{equation}
W\left(  \kappa v\right)  =\int_{0}^{\infty}uS\left(  u\right)  J_{0}\left(
\kappa uv\right)  du\,{\large /}\int_{0}^{\infty}uS\left(  u\right)  du,
\label{VCZ}%
\end{equation}
where $\kappa=2\pi/\left(  \lambda Z\right)  $, with $\lambda$ the wavelength
of light, and $Z$ the distance to the remote screen on which $E$ is measured.
\ Here and throughout $J_{n}$ is the Bessel function of integer order $n$.
\ We define
\begin{equation}
r=\kappa v,
\end{equation}
and hereafter write $W\left(  \kappa v\right)  =W\left(  r\right)  $.

$S\left(  u\right)  $ is the intensity distribution (modulus squared) of the
field leaving the surface of the random medium (sample). \ We assume this
sample to be a random phase plate (diffuser) that is deep enough to yield
Gaussian statistics, but is not so deep as to produce multiple scattering
[$1$]; based on this assumption we equate $S\left(  u\right)  $ with the
intensity distribution illuminating the sample, and assume that the scattered
field retains the state of polarization of the incident field.

\subsection{Simple sources}

We consider initially three different, widely studied \emph{simple} source
distributions that individually generate normal speckle patterns: a Gaussian
(superscript G); a disk (superscript D); and a narrow annulus, or ring,
(superscript R); writing:%
\begin{subequations}
\begin{align}
S^{\left(  \text{G}\right)  }\left(  u\right)   &  =exp\left[  -u^{2}/\left(
4p^{2}\right)  \right]  ;\label{SGauss}\\
S^{\left(  \text{D}\right)  }\left(  u\right)   &  =\Theta\left(  u-p\right)
;\label{SDisk}\\
S^{\left(  \text{R}\right)  }\left(  u\right)   &  =\varepsilon\delta\left(
u-p\right)  , \label{SDelta}%
\end{align}
where $\varepsilon$ in Eq. (\ref{SDelta}) is explained below. Here and
throughout $\Theta\left(  w\right)  $ is the Heaviside step function defined
by $\Theta\left(  w\leq1\right)  =1$, $\Theta\left(  w>1\right)  =0$, and
$\delta\left(  w\right)  $ is the Dirac Delta function.

Although there are no particular problems in generating experimentally the
simple Gaussian and disk source distributions, the ring can only be
approximated. \ This can be done by using a narrow, but finite width, annulus.
\ Writing $\mathbb{S}^{\left(  \text{R}\right)  }\left(  u\right)  $ for this
annulus we have
\end{subequations}
\begin{equation}
\mathbb{S}^{\left(  \text{R}\right)  }\left(  u\right)  =\Theta\left(
u-p-\varepsilon/2\right)  -\Theta\left(  u-p+\varepsilon/2\right)  ,
\label{SR}%
\end{equation}
where $p$ is the mean radius, and $\varepsilon$ the width, of the annulus.
\ Upon passing to the limit of zero width while noting that
\begin{equation}
\lim_{\varepsilon\rightarrow0}\left[  \mathbb{S}^{\left(  \text{R}\right)
}\left(  u\right)  \right]  /\varepsilon=\delta\left(  u-p\right)  ,
\end{equation}
it becomes evident that for consistency with the dimensionless Gaussian and
disk source functions, the source function of the ring must be written as in
Eq. (\ref{SDelta}).

Using Eq. (\ref{VCZ}) the autocorrelation functions $W$ of the simple sources
in Eqs. (\ref{SGauss})-(\ref{SDelta}) are
\begin{subequations}
\label{Wpr}%
\begin{align}
W^{\left(  \text{G}\right)  }\left(  pr\right)   &  =exp\left(  -p^{2}%
r^{2}\right)  ,\label{WG}\\
W^{\left(  \text{D}\right)  }\left(  pr\right)   &  =2J_{1}\left(  pr\right)
/\left(  pr\right)  ,\label{WD}\\
W^{\left(  \text{R}\right)  }\left(  pr\right)   &  =J_{0}\left(  pr\right)  ,
\label{WR}%
\end{align}
whereas the autocorrelation function of the finite width annulus, Eq.
(\ref{SR}), is [$3$]%
\end{subequations}
\begin{subequations}
\begin{align}
\mathbb{W}^{\left(  \text{R}\right)  }\left(  pr\right)   &  =\frac
{1}{p\varepsilon r}[\left(  p+\varepsilon/2\right)  J_{1}\left(
pr+\varepsilon r/2\right) \nonumber\\
&  -\left(  p-\varepsilon/2\right)  J_{1}\left(  pr-\varepsilon r/2\right)
],\label{approxJ0}\\
&  \approx J_{0}\left(  pr\right)  \left[  1-\frac{1}{6}\left(  \frac
{\varepsilon r}{2}\right)  ^{2}\right] \nonumber\\
&  -\frac{\varepsilon^{2}r}{24p}J_{1}\left(  pr\right)  +\mathcal{O}\left(
\left(  \frac{\varepsilon}{2p}\right)  ^{4}\right)  .
\end{align}
As discussed in [$3$], Eq. (\ref{approxJ0}) is an excellent approximation to
Eq. (\ref{WR}) for $\varepsilon r\lesssim1$, so that experiments with narrow
annuli can be expected to closely match the theoretical results obtained here
for a delta function ring.

The inverse of the parameter $p$ is a characteristic length that determines
many properties of the speckle field. \ For example, in all cases the number
density of extrema, singularities, umbilic points, etc. equals $cp^{2}/\pi$,
where the constant $c$, which does not differ very markedly from unity,
depends upon the the form of the autocorrelation function and the nature of
the critical point. \ Thus, the mean separation between critical points is of
order $1/p$, the density of level crossings (line length/unit area) is
proportional to $p$, and the mean spacing between crossings along a straight
line is proportional $1/p$.

\subsection{Composite sources}

These important preliminaries completed, we now turn to composite source
functions that generate speckled speckle fields. \ We build these composite
sources out of pairwise combinations of simple Gaussian, disk, and ring
sources. \ One member of the pair, for which we set $p=a$, has a narrow, but
intense, distribution of intensity, the other member, for which we set $p=b$,
has a broad but weak intensity distribution.

As a specific example we consider a composite source made up out of two
Gaussians. \ We generate this source (conceptually) using a \emph{single}
laser beam and an optical system that produces two concentric foci at the
diffuser - an intense tight focus, the $a$ beam, and a weak broad focus, the
$b$ beam. \ The resulting setup is shown schematically in Fig. \ref{Fig2}.
\ Leaving aside the important, practical details of the optical system that
produces the $a$ and $b$ beams, we proceed to analyze the properties of the
resultant speckled speckle fields.%

\begin{figure}
[ptb]
\begin{center}
\includegraphics[
natheight=8.432800in,
natwidth=6.441100in,
height=2.9793in,
width=2.2822in
]%
{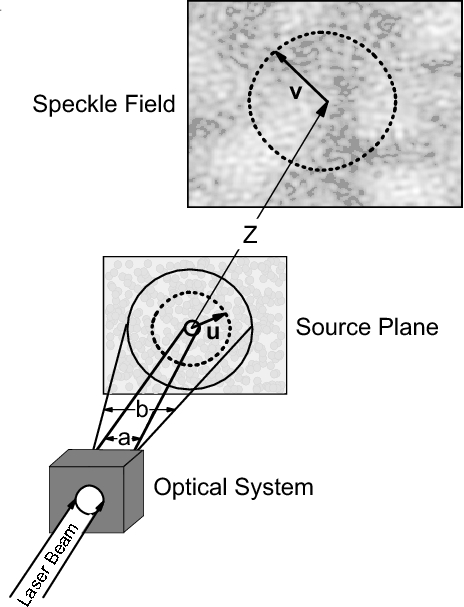}%
\caption{Apparatus for generating speckled speckle. \ A laser beam enters an
optical system that produces two concentric beams which illuminate a deep
phase plate (the source plane). \ One beam, labelled $a$, has a small diameter
and a large intensity, the other, labelled $b$, has a large diameter and a
small intensity. \ The resulting speckle field is observed in the far field of
the source plane. \ Under the paraxial assumption used here, the distance $Z$
between the source plane and speckle field satisfies $Z\gg u,v$. \ The
orthogonal $xy$ axis system in the speckle field can have an arbitrary
orientation, and is not shown. \ The large spots in the speckle field are due
to beam $a$, the small spots to beam $b$. \ For scalar speckled speckle both
beams have the same polarization, for vector speckle speckle they are
orthogonally polarized. \ Using the above geometry, we later illustrate many
of the special properties of speckled speckle with a computer simulation in
which a source that consists of $1000$ randomly phased points (the Source
Plane) with a desired composite amplitude distribution (binary combinations of
Gaussians, Disks, and Rings) radiates into the far field (the Speckle Field).}%
\label{Fig2}%
\end{center}
\end{figure}

We start by noting that now there are two different characteristic lengths,
$\Lambda_{a}=1/a$, and $\Lambda_{b}=1/b$, with $\Lambda_{a}\gg\Lambda_{b}$.
\ In Fig. \ref{Fig1} it is the $a$ field that generates the large speckle
spots, whereas the $b$ field generates the small spots that decorate the $a$
field spots. \ Although this result is, obvious, as becomes apparent there are
a host of nonobvious, often surprising, phenomena that lie hidden in these patterns

Using superscripts T$_{a}$ and T$_{b}$ to denote the source type, where
T$_{a},$T$_{b}\ $= G, D, or R, we write our composite source as%
\end{subequations}
\begin{equation}
S^{\left(  \text{T}_{a}\text{T}_{b}\right)  }\left(  u\right)  =I_{a}%
S^{\text{T}_{a}}\left(  u\right)  +I_{b}S^{\text{T}_{\text{b}}}\left(
u\right)  , \label{Sab}%
\end{equation}
where $I$ is the illuminating intensity (optical power/unit area): for a
Gaussian $I$ is the peak intensity at the center of the Gaussian; for a disk
$I$ is the uniform intensity throughout the disk; and for a ring $I$ is the
uniform intensity within the annulus. \ Inserting Eq. (\ref{Sab}) into Eq.
(\ref{VCZ}) we obtain for the autocorrelation function of the composite source%
\begin{equation}
W^{\left(  \text{T}_{a}\text{T}_{b}\right)  }\left(  r\right)  =\frac
{W^{\left(  \text{T}_{a}\right)  }\left(  ar\right)  +K^{\left(  \text{T}%
_{a}\text{T}_{b}\right)  }W^{\left(  \text{T}_{b}\right)  }\left(  br\right)
}{1+K^{\left(  \text{T}_{a}\text{T}_{b}\right)  }}. \label{Wab}%
\end{equation}
The dimensionless parameter $K^{\left(  \text{T}_{a}\text{T}_{b}\right)  }$ in
Eq. (\ref{Wab}) depends on the combination of sources that are used, and is
given in Appendix A. \ As discussed in this appendix, $K$ equals the total
optical power in the $b$ source beam divided by the total optical power in the
$a$ source beam, so that the composite field autocorrelation function is the
optical power weighted average of the individual $a$ and $b$ autocorrelation functions.

In addition to $W$, we will need throughout later sections mean square
derivatives $r_{2n}$ of the real (imaginary) parts of the composite optical
field, and moments $M_{2n}$ of the composite source functions, in both cases
for $n=1,2,3$;\ $r_{2n}$ is given in Appendix B, $M_{2n}$ in Appendix C. \ We
will also need the dimensionless parameter $\eta=r_{4}/r_{2}^{2}-1$, which is
discussed in Appendix D.

\section{CRITICAL POINTS OF THE PHASE}

\subsection{Scalar Field Statistics and Morphology}

\subsubsection{Statistics of vortices and phase extrema}

\emph{General results.}

The number density $V$ of vortices [$4-6$], $E$ of phase extrema [$7$], and
$S$ of phase saddles, in normal speckle is extended by inspection to speckled
speckle as
\begin{subequations}
\label{pEV}%
\begin{align}
V^{\left(  \text{T}_{a}\text{T}_{b}\right)  }  &  =r_{2}^{\left(  \text{T}%
_{a}\text{T}_{b}\right)  }/\left(  2\pi\right)  ,\label{pEVa}\\
E^{\left(  \text{T}_{a}\text{T}_{b}\right)  }  &  =\frac{\left(
M_{4}^{^{\left(  \text{T}_{a}\text{T}_{b}\right)  }}\right)  ^{3/2}}{4\pi
M_{2}^{^{\left(  \text{T}_{a}\text{T}_{b}\right)  }}\left(  3M_{4}^{^{\left(
\text{T}_{a}\text{T}_{b}\right)  }}-2\left(  M_{2}^{^{\left(  \text{T}%
_{a}\text{T}_{b}\right)  }}\right)  ^{2}\right)  ^{1/2}}\nonumber\\
&  -M_{2}^{^{\left(  \text{T}_{a}\text{T}_{b}\right)  }}/\left(  4\pi\right)
, \label{pEVb}%
\end{align}
where $r_{2}^{\left(  \text{T}_{a}\text{T}_{b}\right)  }$ is given in Appendix
B, $M_{2}^{^{\left(  \text{T}_{a}\text{T}_{b}\right)  }}$ and $M_{4}%
^{^{\left(  \text{T}_{a}\text{T}_{b}\right)  }}$ in Appendix C. \ From the
requirement that the net Poincar\'{e} index of all critical points in the
plane equals $+1$ follows that the densities of vortices, exrema, and saddles,
are related by%
\end{subequations}
\begin{equation}
S=V+E. \label{pSEV}%
\end{equation}

\bigskip

\emph{Statistics for single sources.}

\bigskip

\ $V$ and $E$ for a single source, Eqs. (\ref{SGauss}), (\ref{SDisk}), and
(\ref{SDelta}), are recovered from Eqs. (\ref{pEVa}) and (\ref{pEVb}) by
setting $K=0$, with results [$7$]:
\begin{subequations}
\label{pVGDR}%
\begin{align}
V^{\left(  \text{G}\right)  }  &  =\frac{p^{2}}{\pi},\;\;E^{\left(
\text{G}\right)  }=\frac{p^{2}}{\pi}\left[  \sqrt{2}-1\right]  ,\label{pVG}\\
V^{\left(  \text{D}\right)  }  &  =\frac{p^{2}}{8\pi},\;\;E^{\left(
\text{D}\right)  }=\frac{p^{2}}{8\pi}\left[  2\left(  \frac{2}{3}\right)
^{\left(  3/2\right)  }-1\right]  ,\label{pVD}\\
V^{\left(  \text{R}\right)  }  &  =p^{2}/\left(  4\pi\right)  ,\;\;E^{\left(
\text{R}\right)  }=0. \label{pVR}%
\end{align}
For a single source we therefore have $\left(  E/V\right)  ^{\left(
\text{G}\right)  }=0.414$, $\left(  E/V\right)  ^{\left(  \text{D}\right)
}=0.0887$, $\left(  E/V\right)  ^{\left(  \text{R}\right)  }=0$, so that $E/V$
is always less than unity; for speckled speckle, however, the situation is
remarkably different.

\bigskip

\emph{Anomalous statistics for compound sources.}

\bigskip

In what follows we always take $a<b$, and assume that $\rho=a/b$ is small:
only in this limit are there two widely different length scales in the speckle
pattern. \ We then find that for any given small value of $\rho$ the ratio of
phase extrema to vortices, $\left(  E/V\right)  ^{\left(  \text{T}_{a}%
\text{T}_{b}\right)  }\equiv E^{\left(  \text{T}_{a}\text{T}_{b}\right)
}/V^{\left(  \text{T}_{a}\text{T}_{b}\right)  }$, attains a large, maximum
value, $\left(  E/V\right)  _{max}^{\left(  \text{T}_{a}\text{T}_{b}\right)
}$, when $K^{\left(  \text{T}_{a}\text{T}_{b}\right)  }=K_{max}^{\left(
\text{T}_{a}\text{T}_{b}\right)  }$,where%
\end{subequations}
\begin{align}
K_{max}^{\left(  \text{T}_{a}\text{T}_{b}\right)  }  &  =\frac{m_{2}^{\left(
\text{T}_{a}\right)  }}{m_{2}^{\left(  \text{T}_{b}\right)  }}\rho
^{2},\nonumber\\
\left(  E/V\right)  _{max}^{\left(  \text{T}_{a}\text{T}_{b}\right)  }  &
=\frac{\sqrt{3}m_{4}^{\left(  \text{T}_{b}\right)  }}{12m_{2}^{\left(
\text{T}_{a}\right)  }m_{2}^{\left(  \text{T}_{b}\right)  }}\left(  \frac
{1}{\rho^{2}}\right)  , \label{KEV}%
\end{align}
with $m_{2n}^{\left(  \text{T}\right)  }$ given in Eqs. (\ref{mncn}) and
(\ref{c2n}) . \ For example, for $a$ a Gaussian (G) and $b$ a ring (R),
\begin{subequations}
\label{pKEVmax}%
\begin{align}
K_{max}^{\left(  \text{GR}\right)  }  &  =\frac{m_{2}^{\left(  \text{G}%
\right)  }}{m_{2}^{\left(  \text{R}\right)  }}\rho^{2}=4\rho^{2}%
,\label{pKmax}\\
\left(  E/V\right)  _{max}^{\left(  \text{GR}\right)  }  &  =\frac{\sqrt
{3}m_{4}^{\left(  \text{R}\right)  }}{12m_{2}^{\left(  \text{G}\right)  }%
m_{2}^{\left(  \text{R}\right)  }}\left(  \frac{1}{\rho^{2}}\right)
=\frac{\sqrt{3}}{48\rho^{2}}. \label{pEVmax}%
\end{align}

\bigskip

\emph{Comparison of results for single and compound sources.}

\bigskip

We compare the above results for speckled speckle with those for ordinary
speckle. \ For, say, $\rho=0.1$, which should be achievable in practice
without undue difficulty, we have: for two Gaussians, $\left(  E/V\right)
_{max}^{\left(  \text{GG}\right)  }\approx29$, which is $70$ times larger than
for normal speckle, Eq. (\ref{pVG}); for two disks$\;\left(  E/V\right)
_{max}^{\left(  \text{DD}\right)  }\approx19$, which is $215$ larger than for
normal speckle, Eq. (\ref{pVD}); and for two rings $\left(  E/V\right)
_{max}^{\left(  \text{RR}\right)  }\approx14$. \ This latter result is
especially striking when it is recalled that for normal speckle, Eq.
(\ref{pVR}), there are no phase exrema at all. \ We note that in all three of
the above cases $K_{max}=0.01$, which corresponds to an intensity ratio
$I_{b}/I_{a}=10^{-4}$ and a total optical power ratio $P_{b}/P_{a}=0.01$. \ It
thus seems likely that the major experimental problem in observing the
anomalies of speckled speckle will be the reduction of parasitic light that
would tend to swamp the weak $b$ field intensity $I_{b}$.

\bigskip

\emph{Representative Graphs.}

\bigskip

In Fig. \ref{Fig3} we display graphs of $V$, $E$, and $E/V$, for two Gaussians
(GG), two disks (DD), and two rings (RR). \ The graphs for say DR ($a$ = D,
$b$ = R) or RD ($a$ = R, $b$ = D) can be approximated by dividing the figures
along the line $K=10^{-2}$ and connecting the appropriate combination of
vertically rescaled figure halves;\ for example, the figure for DR is
approximated by attaching the left half of Fig. \ref{Fig3}(b) to the right
half of Fig. \ref{Fig3}(c).%

\begin{figure}
[ptb]
\begin{center}
\includegraphics[
natheight=4.766000in,
natwidth=2.438800in,
height=5.0315in,
width=2.5884in
]%
{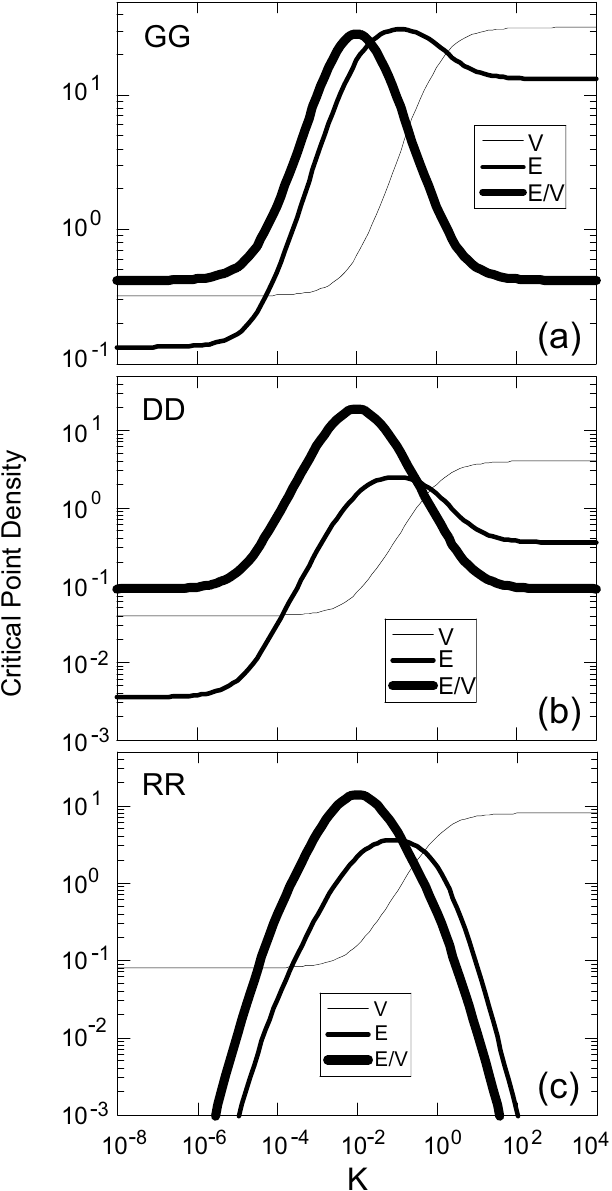}%
\caption{Critical point densities, $V$, $E$, and $E/V$, vs. $K$ for $a=1,b=10$
($\rho=0.1$), for (a) two Gaussians (GG), (b) two disks (DD), and (c) two
rings (RR). \ Note the logarithmic scales.}%
\label{Fig3}%
\end{center}
\end{figure}

The graphs in Fig. \ref{Fig3} are representative, and other source
combinations show similar results: the density of vortices ($V$), Eq.
(\ref{pEVa}), interpolates smoothly from its $a$ value (small $K$) to its $b$
value (large $K$), as does the density (not shown) of saddle points, Eq.
(\ref{pSEV}), whereas both the density of extrema ($E)$, Eq. (\ref{pEVb}), and
the ratio $E/V$, are anomalous, and reach a maximum for small $K$. \ The
density of these maxima can be made arbitrarily large by suitable choice of
$\rho$ and $K$, Eq. (\ref{KEV}).

\subsubsection{Morphology}

\bigskip

\emph{Spatial arrangements of vortices.}

\bigskip

The spatial arrangement of the vortices is also anomalous. \ Because a vortex
is an absolute zero of the wavefunction [$8$], vortices can appear only where
there is complete destructive interference between the strong $a$ and weak $b$
fields, so composite field vortices are constrained to lie in simultaneously
dark regions of the $a$ field, and the composite field, speckle patterns, Fig.
\ref{Fig4}. \ For very small $I_{b}/I_{a}$ the $b$ field is a minor
perturbation of the $a$ field. In this limit the sparse zeros, and therefore
the sparse vortices, of the $a$ field maintain their identities but are
shifted slightly in position, Fig. \ref{Fig4}(a). For larger intensity ratios,
however, both $a$ and $b$ field vortices lose their identities.%

\begin{figure}
[pb]
\begin{center}
\includegraphics[
natheight=6.319200in,
natwidth=3.693600in,
height=2.8712in,
width=1.689in
]%
{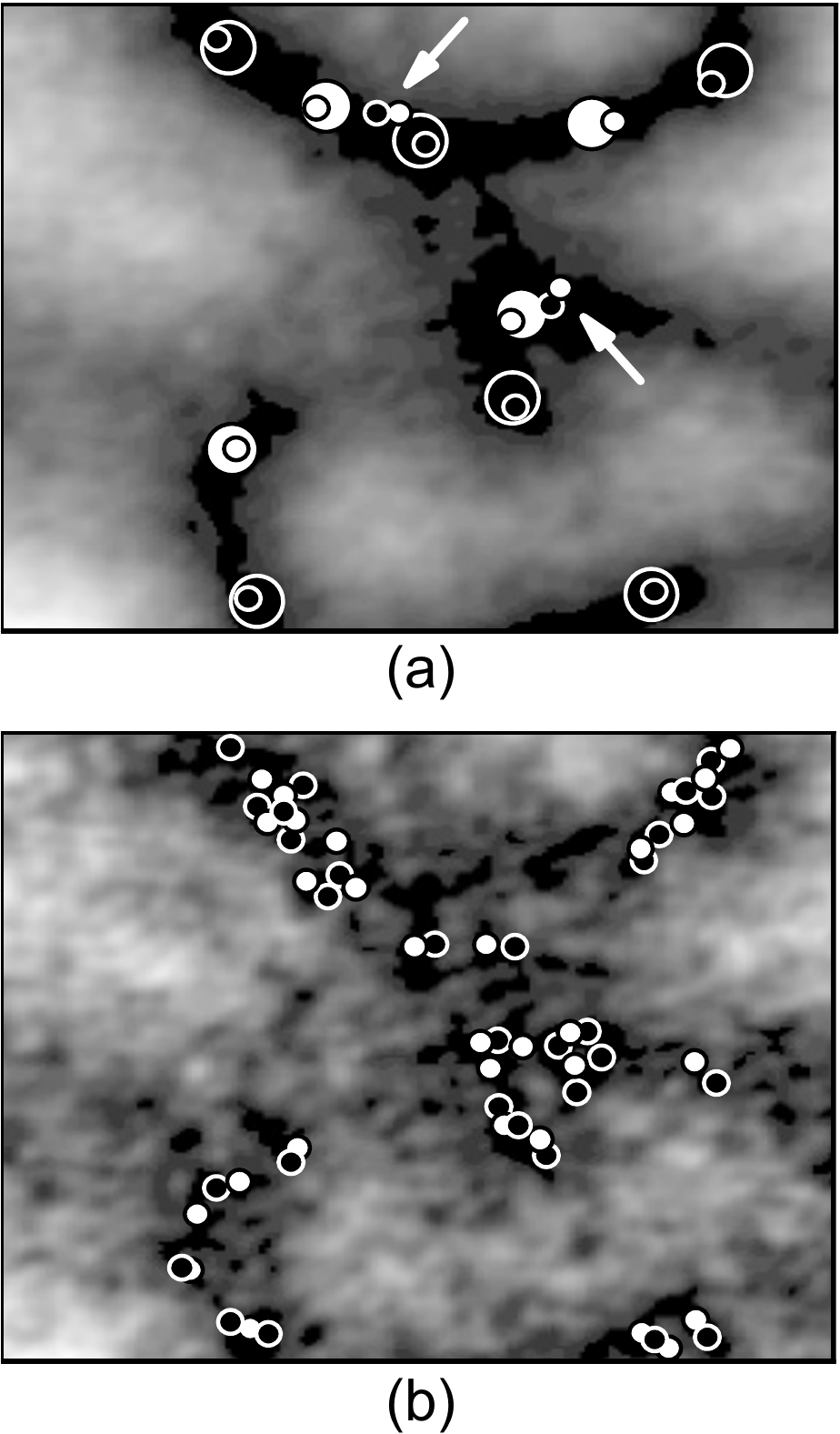}%
\caption{Speckled speckle phase vortices for Gaussian $a$ and $b$ fields (GG)
with $\rho=a/b=0.1$. \ Shown is a typical region of the intensity speckle
pattern with vortices superimposed, where filled white circles with black rims
(filled black circles with white rims) represent positive (negative) vortices.
\ (a) $I_{b}/I_{a}=2\times10^{-5}$ ($K=0.002$). \ The large circles show the
positions of the vortices of the strong $a$ field ($I_{b}=0$), the small
circles show the vortices of the composite field. \ As expected, for the small
$I_{b}/I_{a}$ ratio used here, the composite field vortices cluster near the
positions of the sparse $a$ field vortices, which are the darkest regions of
the speckle pattern.\ \ As $I_{b}/I_{a}$ is increased from zero, and as is
illustrated here, initially the $a$ field vortices undergo small displacements
but maintain their identities, while a small number of new positive/negative
vortex pairs nucleate (here two pairs, arrows). \ (b) \ The field in (a) for
$I_{b}/I_{a}=10^{-3}$ ($K=0.1$). \ As $I_{b}/I_{a}$ increases the number of
composite field vortices increases rapidly (Fig. \ref{Fig3}), and in addition
to vortex nucleation vortex annihilation also occurs, leading to a loss of
identity of the $a$ field vortices. \ For the moderately small $I_{b}/I_{a}$
ratio used here the composite field vortices no longer need cluster around the
positions of the $a$ field vortices; they are, however, still constrained, to
lie in the dark regions surrounding the large $a$ field speckle spots. \ For
still larger $K$ vortices begin to fill the field more uniformly, and in the
limit of large $K$ the dense vortex structure is that of the $b$ field. }%
\label{Fig4}%
\end{center}
\end{figure}

\bigskip

\emph{Reactions between vortices and phase stationary points.}

\bigskip

As $I_{b}$ increases new vortices appear. \ Conservation of charge requires
these vortices to nucleate as positive/negative pairs. \ Pair production by
itself, however, violates conservation of the Poincar\'{e} index, here $I_{P}%
$, because $I_{P}=+1$ for vortices independent of their charge. \ Vortex pair
production must therefore be accompanied by another process, and in general
there are two possibilities:

\bigskip

I. A vortex pair and a pair of saddle points (each one of which has
$I_{P}=-1)$ are generated [$9$].

\medskip

II. Two extrema (one minimum and one maximum, each of which has $I_{P}=+1$)
collide and annihilate, and a vortex pair emerges from the collision [$10$].

\bigskip

For very small $I_{b}$ there are, in general, fewer extrema than vortices, so
in this regime mechanism II cannot be operative (recall that for a ring there
are no phase extrema at all), leaving mechanism I as the only alternative.
\ In Fig. \ref{Fig3} this regime corresponds to the region of say $K<10^{-4}$.
\ But as can be seen from this figure, for larger $K$ the number of extrema
exceeds the number of vortices, thereby permitting mechanism II to become
operative. \ These two reaction mechanisms are illustrated in Fig. \ref{Fig5}
for the representative case of GG for small and intermediate $K$, and in Fig.
\ref{Fig6} for RR for larger $K$; as can be seen from these figures, for
sufficiently small (large) $K$ mechanism I (mechanism II) dominates.

\begin{figure}
[pt]
\begin{center}
\includegraphics[
natheight=4.585200in,
natwidth=3.986800in,
height=2.5495in,
width=2.2208in
]%
{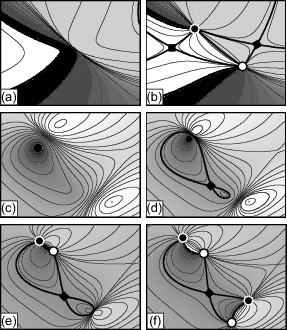}%
\caption{Phase maps showing vortex nucleation for small and intermediate $K$
for Gaussian $a$ and $b$ source fields (GG) with $a/b=0.1$. \ The phase
increases $0$ to $2\pi$ black to white. \ Filled black (white) circles
represent negative (positive) vortices, whereas black squares mark saddle
points. \ Thin curves are ordinary contours, thick curves are the special
contours (bifurcation lines) that pass through the saddle points. \ (a),(b)
Vortex nucleation for very small $K$. \ (a) \ $K=0.00164$. \ (b) $K=0.0017$.
\ In this small $K$\ region mechanism I of the text is dominant, and vortex
pairs nucleate together with a pair of saddle points. \ The resulting box like
structure, first described in [$9$], is the one usually observed following
vortex pair production in normal speckle. \ (c)-(f) Intermediate $K$ regime.
\ In (c), (d), (e), and (f), $K=0.024,0.0275,0.0291,$ and $0.0325$,
respectively. \ In this region mechanism II of the text\ dominates.
\ Specifically: in (c) there are no vortices, only phase extrema (two maxima
and one minimum); in (d) the upper minimum and maximum approach one another,
while in an unrelated process a new saddle point and a new minimum nucleate;
in (e) the upper pair of extrema collide and generate a positive/negative
vortex pair; and in (f) the lower pair of extrema collide and generate a
second vortex pair.}%
\label{Fig5}%
\end{center}
\end{figure}

%

\begin{figure}
[pb]
\begin{center}
\includegraphics[
natheight=4.760800in,
natwidth=2.763900in,
height=2.6455in,
width=1.548in
]%
{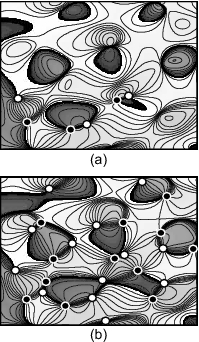}%
\caption{Phase maps showing vortex nucleation for larger values of $K$ for
both $a$ and $b$ source fields rings (RR) with $a/b=0.1$. \ Symbols are as in
Fig. \ref{Fig5}. \ (a) $K=0.5$. \ (b) $K=2$. \ In (a), which corresponds
approximately to the region of the flat maximum in the number of extrema E in
Fig. \ref{Fig3}c, there are three vortex pairs and numerous maxima and minima.
\ In (b), which corresponds to the onset of rapid decline in the number of
extrema in Fig. \ref{Fig3}c, all maxima and minima in (a) have collided,
annihilated, and been replaced by vortex pairs. \ }%
\label{Fig6}%
\end{center}
\end{figure}

\subsection{Vector Field C points and Extrema}

\subsubsection{Introductory remarks}

The state of polarization of generic vector fields is elliptical. In the
paraxial fields of interest here the (always planar [$11$]) polarization
ellipses lie in planes, here the $xy$-plane, oriented normal to the direction
of propagation, the $Z$-axis. \ The point singularities of such fields in the
plane (in space) are C points (form continuous C lines), which are points
(lines) of circular polarization embedded in a field of elliptical
polarization [$12$]. \ Because the azimuthal orientation of a circle (the C
circle) is undefined, the surrounding ellipses rotate about the point,
generically with rotation (winding) angle $\pm\pi$, and topological charge
(winding number) $\pm%
\frac12
$ for positive/negative points [$12$].

Necessarily associated with C points in the planes of generic paraxial fields
are azimuthal stationary points - extrema (maxima and minima%
$\vert$%
), and saddle points [$13$]. \ Unlike C points which are easily identified
visually as the centers around which the surrounding ellipses rotate,
azimuthal stationary points are subtle features that may be difficult to
discern in traditional maps that show polarization ellipses. \ A highly useful
representation that shows both types of critical points clearly is the phase
(argument) $\Phi_{12}$ of the complex Stokes field [$14$]%
\end{subequations}
\begin{equation}
S_{12}=S_{1}+i\,S_{2} \label{Stokes12}%
\end{equation}
\ In this representation C points are vortices with charges $\pm1$, and
azimuthal stationary points are phase maxima, minima, and saddle points. \ In
what follows we use this representation exclusively.

Charge conservation requires that C points nucleate or annihilate as
positive/negative pairs, whereas conservation of the Poincar\'{e} index
$I_{P}$, which is $+1$ for C points of either sign, requires that pair
production be accompanied by either the production of two azimuthal saddles,
for which $I_{P}=-1$, or the disappearance of two extrema, for which
$I_{P}=+1$.

Like ordinary circularly polarized light, C points are either right handed or
left handed, depending upon the direction of rotation (clockwise or
counterclockwise) of the electric vector of the light as it traces out the C
circle. \ By continuity, the polarization ellipses in the immediate vicinity
of a C point have the handedness of the point, so that overall the ellipse
field is divided into right and left handed regions. \ Near the boundary
between regions of opposite handedness the polarization ellipses narrow, and
on the boundary itself the ellipses collapse into lines and the polarization
is linear; these lines are called L lines [$12$].

All of the above holds for both ordinary, and speckled speckle, vector fields.

We now turn to a discussion of the number densities of C points and their
associated extrema in vector speckled speckle. \ Here the $a$ and $b$ source
fields have orthogonal polarizations, and we consider two different cases: in
the first the $a$ ($b$) source field is linearly polarized along the $x$-axis
($y$-axis); in the second the $a$ ($b$) source field has right (left) circular
polarization. \ As becomes apparent, these two different configurations
produce very different results.

\subsubsection{Linearly polarized source fields: statistics and ellipse field
morphology}

\bigskip

\hspace{-0.15in}\emph{a. \ C point densities.}

\bigskip

In what follows we assume, as already stated, that speckle field $E_{x}$
($E_{y}$) maintains the state of linear polarization of its $a$ ($b$) source.
We begin by noting that at a C point the amplitudes of $E_{x}$ and $E_{y}$
must be equal, so like phase vortices, C points of the composite field are
found in the dark regions of the $a$ field speckle pattern [$2$]. \ But unlike
vortices, these regions need not also be dark regions of the composite field
speckle pattern, because unlike a vortex, a C point is not a zero of the
vector field.

In considering number densities it is convenient to decompose the
\emph{speckle} field, Fig. \ref{Fig2}, into right (R) and left (L) circularly
polarized components, $E_{R}$ and$\ E_{L}$. \ The reason is that $E_{R}$
($E_{L}$) vanishes at a right handed (left handed) C point, and since the
single polarization component fields $E_{R}$ and $E_{L}$ are scalar fields,
Eq. (\ref{pEVa}) is applicable to each component. \ From this follows that in
normal speckle fields the number density of C points is twice the density of
phase vortices [$12$].

Eq. (\ref{pEVa}) is also applicable to C points in vector speckled speckle,
but to use this result we need derivatives $r_{2}^{\left(  \text{T}%
_{a}\text{T}_{b}\right)  }$ for the \emph{circularly} polarized components.
\ Using the results in Appendix B, we can obtain these derivatives from the
autocorrelation functions $W_{R,L}^{\left(  \text{T}_{a}\text{T}_{b}\right)
}$ of $E_{R}$ and$\ E_{L}$; these autocorrelation functions are calculated below.

We start by writing the circularly polarized components $E_{R}$ \ and $E_{L}$
in terms of the Cartesian components $E_{x}$ and $E_{y}$ of the speckle field
as%
\begin{equation}
E_{R,L}=\left(  E_{x}\mp iE_{y}\right)  /\sqrt{2}, \label{ERL}%
\end{equation}
where $E_{x}$ ($E_{y}$) arises from the $x$-axis ($y$-axis) linearly polarized
$a$ ($b$) source field. \ Temporarily suppressing for notational convenience
superscripts T$_{a}$T$_{b}$, $W_{R,L}$ is%
\begin{equation}
W_{R,L}\left(  r\right)  =\left\langle E_{R,L}^{\ast}\left(  0\right)
E_{R,L}\left(  r\right)  \right\rangle /\left\langle I_{R,L}\right\rangle
\end{equation}
where the average intensity $\left\langle I_{R,L}\right\rangle =\left\langle
E_{R,L}^{\ast}\left(  0\right)  E_{R,L}\left(  0\right)  \right\rangle
=\tfrac{1}{2}\left(  \left\langle I_{x}\right\rangle +\left\langle
I_{y}\right\rangle \right)  $, with $\left\langle I_{x}\right\rangle $ and
$\left\langle I_{y}\right\rangle $ the average intensities of the Cartesian
components. \ Using Eq. (\ref{ERL}) together with $\left\langle E_{x}^{\ast
}\left(  0\right)  E_{y}\left(  r\right)  \right\rangle =\left\langle
E_{y}^{\ast}\left(  0\right)  E_{x}\left(  r\right)  \right\rangle
=\left\langle E_{x}^{\ast}\left(  0\right)  E_{y}\left(  0\right)
\right\rangle =\left\langle E_{y}^{\ast}\left(  0\right)  E_{x}\left(
0\right)  \right\rangle =0$, we have%
\begin{equation}
W_{R}^{\left(  \text{T}_{a}\text{T}_{b}\right)  }\left(  r\right)
=W_{L}^{\left(  \text{T}_{a}\text{T}_{b}\right)  }\left(  r\right)
=W^{\left(  \text{T}_{a}\text{T}_{b}\right)  }\left(  r\right)  , \label{WRL}%
\end{equation}
where $W^{\left(  \text{T}_{a}\text{T}_{b}\right)  }\left(  r\right)  $ is
given in Eq. (\ref{Wab}), and $K$ is given in Eq. (\ref{KPbPa}). \ In
obtaining this result we use the fact that$\ \left\langle I_{x}\right\rangle $
($\left\langle I_{y}\right\rangle $) is proportional to the total optical
power in the source field $a$\ ($b$) component, so that $\left\langle
I_{y}\right\rangle /\left\langle I_{x}\right\rangle =K$. \ Thus, Eq.
(\ref{pEVa}) and curves $V$ in Fig. \ref{Fig3}, multiplied by a factor of two,
also describe C points in vector speckled speckle.

\bigskip

\hspace{-0.15in}\emph{b. \ L line densities.}

\bigskip

The density (line length/unit area) of L lines is proportional to the square
root of the density of C points [$15$]. \ Because the density of C points,
$2V$, interpolates smoothly between its small $K$ ($a$ field) and large $K$
($b$ field) limits, so does the L line density, and L lines in speckled
speckle do not show anomalies.

\bigskip

\hspace{-0.15in}\emph{c. \ Densities of azimuthal extrema.}

\bigskip

\emph{K and }$\rho=a/b$\emph{ both small.}

\bigskip

Although Eq. (\ref{pEVa}) is applicable to speckled speckle C points, Eq.
(\ref{pEVb}) is not necessarily applicable to the corresponding azimuthal
stationary points; the reason is that unlike the case of C points, there is no
a priori simple relationship between vector field azimuthal stationary points
and the stationary points of the corresponding scalar field components. \ At
present there is no exact theory for the azimuthal densities; we are, however,
able to estimate these densities when both $K$ and $a/b$ are small.

Returning to Eq. (\ref{ERL}) we write%
\begin{equation}
\sqrt{2}E_{R,L}=E_{x}\left(  1\mp iE_{y}/E_{x}\right)  , \label{eReL}%
\end{equation}
where the motivation for this form is the fact that when $K$ is small so also
is $E_{y}/E_{x}$. \ Writing the various field components in terms of
amplitudes $A$ and phases $\varphi$ or $\Phi$,
\begin{subequations}
\label{ExEyErEl}%
\begin{align}
E_{x}  &  =A_{x}\exp\left(  i\varphi_{x}\right)  ,\label{Ex}\\
E_{y}  &  =A_{y}\exp\left(  i\varphi_{y}\right)  ,\label{Ey}\\
E_{R}  &  =A_{R}\exp\left(  i\Phi_{R}\right)  ,\label{Er}\\
E_{L}  &  =A_{L}\exp\left(  i\Phi_{L}\right)  , \label{El}%
\end{align}
and noting that Stokes parameters $S_{1}$ and $S_{2}$ can be written%
\end{subequations}
\begin{subequations}
\begin{align}
S_{1}  &  =2\operatorname{Re}\left(  E_{R}^{\ast}E_{L}\right)  ,\label{S1}\\
S_{2}  &  =2\operatorname{Im}\left(  E_{R}^{\ast}E_{L}\right)  , \label{S2}%
\end{align}
we have from Eq. (\ref{Stokes12})
\end{subequations}
\begin{equation}
S_{12}=2E_{R}^{\ast}E_{L}, \label{S12}%
\end{equation}
from which follows%
\begin{equation}
\Phi_{12}=\Phi_{L}-\Phi_{R}. \label{PHI12}%
\end{equation}

Returning to Eq. (\ref{eReL}) we have to third order in the small quantity
$E_{y}/E_{x}$,%
\begin{align}
1\pm iE_{y}/E_{x}  &  \approx\exp\left(  (E_{y}/E_{x})^{2}/2\right)
\nonumber\\
&  \times\exp\left(  \mp\text{i\thinspace}\left(  E_{y}/E_{x}-\left(
E_{y}/E_{x}\right)  ^{3}/3\right)  \right)  . \label{smallExEy}%
\end{align}
Writing%
\begin{equation}
E_{y}/E_{x}=\left(  A_{y}/A_{x}\right)  \left[  \cos\left(  \varphi
_{y}-\varphi_{x}\right)  +\text{i\thinspace}\cos\left(  \varphi_{y}%
-\varphi_{x}\right)  \right]  , \label{EyEx}%
\end{equation}
we have%
\begin{align}
\Phi_{R,L}  &  =\varphi_{x}\mp\left(  A_{y}/A_{x}\right)  \cos\left(
\varphi_{y}-\varphi_{x}\right) \nonumber\\
&  +\tfrac{1}{2}(A_{y}/A_{x})^{2}\sin\left(  2\left(  \varphi_{y}-\varphi
_{x}\right)  \right) \nonumber\\
&  \pm\tfrac{1}{3}(A_{y}/A_{x})^{3}\cos^{3}\left(  \varphi_{y}-\varphi
_{x}\right) \nonumber\\
&  \mp(A_{y}/A_{x})^{3}\sin\left(  \varphi_{y}-\varphi_{x}\right)  \sin\left(
2\left(  \varphi_{y}-\varphi_{x}\right)  \right)  . \label{PhyRL}%
\end{align}
so that to leading order in $A_{y}/A_{x}$,
\begin{subequations}
\label{Phy12abc}%
\begin{align}
\Phi_{12}  &  \approx-2\left(  A_{y}/A_{x}\right)  \cos\left(  \varphi
_{y}-\varphi_{x}\right) \label{Phy12a}\\
&  \approx2\left(  \Phi_{L}-\varphi_{x}\right) \label{Phy12b}\\
&  \approx-2\left(  \Phi_{R}-\varphi_{x}\right)  . \label{Phy12c}%
\end{align}
Eqs. (\ref{Phy12abc}) make a number of possibly unexpected predictions that
are listed below.

\bigskip

\emph{Density of azimuthal extrema for }$K$\emph{ equal to its value at the
peak of curve E in Fig. \ref{Fig3}.}

\bigskip

Far from its sparse vortices the phase $\varphi_{x}$ of the slowly varying $a$
field changes slowly, so that in regions where $\Phi_{R,L}$ vary rapidly,
$\varphi_{x}$ in Eqs. (\ref{Phy12b}) and (\ref{Phy12c}) can be treated locally
as an unimportant, nearly flat background that can be eliminated by a local
redefinition of the zero of phase. \ For sufficiently small $\rho=a/b$ these
flat regions occupy most of the wavefield area and we can write%
\end{subequations}
\begin{equation}
\Phi_{12}\approx2\Phi_{L}\approx-2\Phi_{R}. \label{Phy12RL}%
\end{equation}
In the region of the peak of curve E in Fig. \ref{Fig3}, where there is a high
density of extrema, $\Phi_{R,L}$ vary rapidly. \ We therefore conclude that
near this peak the density of azimuthal extrema very nearly equals the density
of phase extrema $E$, so that just like for scalar speckle speckle, $E/V$ for
vector speckle speckle can attain very large values.

\begin{figure}
[pb]
\begin{center}
\includegraphics[
natheight=3.007000in,
natwidth=3.976400in,
height=1.8317in,
width=2.4128in
]%
{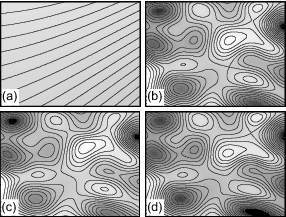}%
\caption{Azimuthal extrema in vector speckled speckle for $a/b=0.01$, and
$K=10^{-4}$. \ The amplitude in these contour maps increases black to white,
so that high maxima (deep minima) are colored white (black). \ (a)
$\varphi_{x}$. The variation in $\varphi_{x}$ is less than $7\%$ of the
variation of $\Phi_{12}$ over this same region, so to first order $\varphi
_{x}$ can be taken to be constant in Eqs. (\ref{Phy12abc}). \ (b) $\Phi_{12}$.
\ (c) $-\Phi_{R}$. \ The minus sign turns maxima into minima and vice versa,
facilitating comparison with $\Phi_{12}$ and $\Phi_{L}$. \ (d) $\Phi_{L}$.
\ Note the close similarity of the three fields $\Phi_{12}$, $\Phi_{L}$, and
$\Phi_{L}$, in full accord with Eqs. (\ref{Phy12RL}). }%
\label{Fig7}%
\end{center}
\end{figure}

In Fig. \ref{Fig7} we compare $\varphi_{x}$, $\Phi_{12}$, $\Phi_{L}$, and
$-\Phi_{R}$ for the case of two disks (DD) for the small, but still
experimentally possible, ratio $a/b=0.01$, for $K=1\times10^{-4}$, which
corresponds to the maximum of $E/V$ for the above $a/b$ ratio (here $\left(
E/V\right)  _{\max}=1924$). \ These parameters are favorable ones, and for
larger $a/b$ ratios, say $a/b=0.1$, the very close agreement with the
predictions of Eqs. (\ref{Phy12RL}) degrades, because now $\varphi_{x}$ can no
longer be considered to be locally flat over most of the wavefield area; the
overall conclusion, however, remains, and in general $E/V$ is anomalously
large for vector speckle speckle.%

\bigskip

\emph{Density of azimuthal extrema for }$K$\emph{ less than its value at the
peak of curve E in Fig. \ref{Fig3}.}

\bigskip

\begin{figure}
[pb]
\begin{center}
\includegraphics[
natheight=1.669100in,
natwidth=3.976400in,
height=2.4699in,
width=2.4111in
]%
{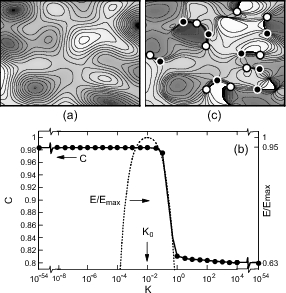}%
\caption{Azimuthal extrema for two disks (DD) for $a/b=0.01$, as in Fig.
\ref{Fig7}. (a) $\Phi_{12}$ for $K=10^{-54}$. \ Note that the stationary point
structure here is very nearly the same as in Fig. \ref{Fig7}(b) where
$K=10^{-4}$ is fifty orders of magnitude larger. \ (b) Correlation coefficient
$C=C\left(  \Phi_{L}\left(  K_{0}\right)  ,\Phi_{12}\left(  K\right)  \right)
$, Eq. (\ref{CCoef}). $\ \Phi_{L}$ is the phase of the left circularly
polarized vector field component, $\Phi_{12}$ is the Stokes phase of the
combined vector field. \ $K_{0}=$ $10^{-2}$ is the value of $K$ at which the
scaled number density $E/E_{\max}$ of $\Phi_{L}$ extrema, dotted curve,
attains its maximum value. Curve\ $E/E_{\max}$ is plotted as $1+\log
_{10}\left(  E/E_{\max}\right)  $, so that where $C=0.98$ ($C=0.8$),
$E/E_{\max}=$ $0.95$ ($E/E_{\max}=$ $0.63$). \ Note the extreme range of $K$,
$10^{-54}\leq K\leq$ $10^{54}$. \ \ (c) $\Phi_{12}$ for $K=1$. \ As can be
seen, when $K$ starts to become large the azimuthal maxima and minima in (a)
start to collide and annihilate, generating positive/negative C point pairs
(filled white/black circles). \ }%
\label{Fig8}%
\end{center}
\end{figure}

For small ratios $a/b$ for all values of $K$ less than $K_{0}$, the value of
$K$ at the peak of the $E$ curve, the stationary point structure of $\Phi
_{12}$ remains very nearly independent of $K$. \ This follows from the fact
that not only the phase $\varphi_{x}$ of the slowly varying $a$ field hardly
changes, but that also $A_{x}$, the amplitude of the field, changes slowly,
and can therefore locally be taken to be constant. \ Eliminating $\varphi_{x}$
by redefining the local zero of phase, and replacing $A_{x}$ by a local
constant $c$, we have%
\begin{equation}
\Phi_{12}\approx-cA_{y}\cos\left(  \varphi_{y}\right)  . \label{Phy12y}%
\end{equation}
Now, for a given value of $\rho$,\ the structure of $A_{y}$, and the structure
and magnitude of $\varphi_{y}$, are independent of $I_{a}$ and $I_{b}$, and
are therefore independent of $K$; Eq. (\ref{Phy12y}) therefore implies that
although the magnitude of $\Phi_{12}$ changes with $K$, its structure remains
invariant. \ This is illustrated qualitatively in Fig. \ref{Fig8}(a).

In Fig. \ref{Fig8}(b) we quantify the relationship between $\Phi_{12}$ and
$\Phi_{L}$. \ Shown is the cross correlation coefficient $C$ between
$\Phi_{12}\left(  K\right)  $ and $\Phi_{L}\left(  K_{0}\right)  $, where
$C=C\left(  u,v\right)  $ for two real variables $u,v$ is%
\begin{equation}
C\left(  u,v\right)  =\frac{\left\langle \left(  u-\left\langle u\right\rangle
\right)  \left(  v-\left\langle v\right\rangle \right)  \right\rangle }%
{\sqrt{\left\langle \left(  u-\left\langle u\right\rangle \right)
^{2}\right\rangle \left\langle \left(  v-\left\langle v\right\rangle \right)
^{2}\right\rangle }}. \label{CCoef}%
\end{equation}
Here $\left\langle ...\right\rangle $ represents spatial averages over the
phase fields. As can be seen, $\Phi_{12}$ and $\Phi_{L}$ are nearly identical
($C=0.983$) up to and including the point at which $E$ passes through its
maximum, showing that in vector speckled speckle the density of azimuthal
extrema all throughout the small $K$ region substantially equals the
\emph{maximum} value of $E$ for scalar speckled speckle.

An interesting, possibly counterintuitive, feature of Fig. \ref{Fig8}(b) is
the significant degree of correlation ($C=0.8$) between $\Phi_{12}$ and
$\Phi_{L}$ in the large $K$ region where most extrema have been converted to C
points. \ This correlation reflects the fact that locally the phase field
before and after collisions of extrema maintains more-or-less the same value.
\ This behavior is most clearly seen in Fig. \ref{Fig6}, where dark (light)
regions in Fig. \ref{Fig6}(a) for the most part remain dark (light) in Fig.
\ref{Fig6}(b).

\bigskip

\emph{Generation of C points as }$K$\emph{ increases from zero.}

\bigskip

As noted above, the density of C points increases with $K$ in accord with
curves $V$ (multiplied by two) in Fig. \ref{Fig3}. \ For scalar speckled
speckle vortex generation in the small $K$ region in which extrema are sparse
occurs via mechanism I in which a new vortex pair appears together with a pair
of saddle points, Figs. \ref{Fig5}(a),(b), whereas in the region of larger $K$
in which extrema are plentiful vortex pairs are generated via mechanism II in
which the collision and annihilation of a pair of extrema, one maximum and one
minimum, generates the vortices, Figs. \ref{Fig5}(c)-(f) and Fig. \ref{Fig6}.
\ For C points, however, extrema are abundant for both small and large $K$,
and it is mechanism II, i.e. collision of azimuthal extrema, that dominates C
point production for all $K$.%

\bigskip

\hspace{-0.15in}\emph{d. \ Decrease in the number of azimuthal extrema for
large }$K$\emph{.}

\bigskip

In the limit of large $K$ where $E_{x}/E_{y}$ is small, the analog of Eq.
(\ref{eReL}) is%
\begin{equation}
\sqrt{2}E_{R,L}=\mp\text{i\thinspace}E_{y}\left(  1\pm\text{i\thinspace}%
E_{x}/E_{y}\right)  , \label{eYeX}%
\end{equation}
and the analogs of Eqs. (\ref{Phy12abc}) are
\begin{subequations}
\label{largeKabc}%
\begin{align}
\Phi_{12}  &  \approx-2\left(  A_{x}/A_{y}\right)  \cos\left(  \varphi
_{x}-\varphi_{y}\right)  +\pi,\label{largeKa}\\
&  \approx2\left(  \Phi_{L}-\varphi_{y}\right)  ,\label{largeKb}\\
&  \approx-2\left(  \Phi_{R}-\varphi_{y}\right)  . \label{largeKc}%
\end{align}
For large $K$, in scalar fields $\Phi_{L}$, $\Phi_{R}$, and $\varphi_{y}$, $E$
decreases from its large maximum to its small single component value (Fig.
\ref{Fig3}). \ Eqs. (\ref{largeKabc}) therefore imply that also the large
density of extrema present in vector speckled speckle for small $K$ decreases
for large $K$. \ For scalar fields the reduction in the number of extrema
results from collisions of maxima and minima that produce vortex pairs, Figs.
\ref{Fig5} and \ref{Fig6}; as is illustrated in Fig. \ref{Fig8}(c), the same
is true for vector speckled speckle.

\bigskip

\hspace{-0.15in}\emph{e. C points for vanishingly small values of either the
}$a$\emph{ or the }$b$\emph{ field.}

\bigskip

As $K$ increases the phase structure stabilizes asymptotically at a mix of C
points and extrema, with C points outnumbering extrema by perhaps a factor of
$3-5$. \ But if for, say, $\left\vert E_{x}\right\vert =10^{-25}\left\vert
E_{y}\right\vert $ ($K=10^{54}$), how, for all intents and purposes, can the
wavefield be anything other than linearly polarized? \ The answer [$12$, sect.
12.5] is that $E_{y}$ goes continuously to zero at its vortices, so that
surrounding each $E_{y}$ vortex, which is an $I_{y}=\left\vert E_{y}%
\right\vert ^{2}$ zero minimum, there is for every $K$ an elliptical contour
on which $\left\vert E_{x}\right\vert =\left\vert E_{y}\right\vert $. The size
of this contour shrinks with increasing $K$, so for sufficiently large $K$ the
phase $\varphi_{x}$ of $E_{x}$ on the contour can be taken to be constant.
\ Because the phase field surrounding the $E_{y}$ vortex winds through $2\pi$
there are always two points on the contour (separated by $180^{o}$) where
$\left\vert \varphi_{y}-\varphi_{x}\right\vert =\pi/2$; at these points a
positive/negative pair of C points appears. \ Similar considerations apply to
the limit of very small $K$, where the $C$ points are located near the
vortices of $E_{x}$.

\bigskip

\hspace{-0.15in}\emph{f. \ Comparison of results for scalar and vector
fields.}

\bigskip

We conclude this subsection with a summary of the similarities and differences
in the number densities of vortices, C points, and extrema, in scalar and in
linearly polarized speckled speckle: \ (\emph{i}) For both cases the density
of singularities - vortices for the scalar case, C points for the linearly
polarized vector case - interpolates smoothly with $K$ from its small $a$
field value to its large $b$ field value: Fig. \ref{Fig3}, curve $V$ for
vortices; this same curve multiplied by two for C points. \ (\emph{ii}) For
the scalar case the density of extrema, and the ratio of extrema to vortices,
increase from their $a$ field values to large maxima at small $K$, and then
decrease for large $K$ to their $b$ field values: Fig. \ref{Fig3}; Eqs.
(\ref{KEV}). \ (\emph{iii}) For the vector case the density of extrema, and
the ratio of extrema to C points, are large for all small $K$, and decrease
for large $K$. \ For sufficiently small $a/b$ the large azimuthal extrema
density for small $K$ substantially equals the maximum density of phase
extrema for the scalar case, whereas the large ratio of azimuthal extrema to C
points equals half the maximum ratio of phase extrema to vortices for the
scalar case. \ (\emph{iv}) In the scalar case generation of vortex pairs is
accompanied by nucleation (annihilation) of a pair of saddle points (extrema)
for small (large) $K$. \ For the vector case C point generation in speckled
speckle occurs through the annihilation of paired azimuthal extrema for all
$K$. \ (\emph{v}) The reduction in the number of extrema for large $K$ results
from collisions of maxima and minima to produce vortices in the scalar case, C
points in the vector case.

\subsubsection{Circularly polarized source fields}

We take the $a$ ($b$) field to be right (left) circularly polarized, and write
the circular components in the speckle plane as
\end{subequations}
\begin{subequations}
\label{EraELb}%
\begin{align}
E_{R}  &  =A_{a}\exp\left(  i\varphi_{a}\right)  ,\label{ERa}\\
E_{L}  &  =A_{b}\exp\left(  i\varphi_{b}\right)  . \label{ELb}%
\end{align}
Eq. (\ref{PHI12}) then becomes%
\end{subequations}
\begin{equation}
\Phi_{12}=\varphi_{b}-\varphi_{a}, \label{PHIba}%
\end{equation}
which implies the following: (\emph{i}) The number densities, and the spatial
distributions of all critical points - C points, azimuthal extrema, and
azimuthal saddle points - are independent of $K$, which depends on the
amplitudes of the fields, but not on their phases. \ (\emph{ii}) For small
values of $a/b$ for which the slowly varying $a$ field phase $\varphi_{a}$
can, over most of the wavefield area, be treated locally as a constant, the
critical point structure of $\Phi_{12}$ is essentially the same as that of
$\varphi_{b}$. \ In this limit the critical point number densities are those
of scalar field $b$, and are given by Eqs. (\ref{pVGDR}) with $p=b$. \ These
conclusions are fully verified by our computer simulations (not shown).
\ Thus, in contrast to linearly polarized speckled speckle with its rich set
of anomalies, circularly polarized speckled speckle appears to offer little
that is of special interest.

\subsection{Zero Level Crossings of Phase Derivatives}

\medskip

\emph{Introductory remarks}.

\medskip

Nodal domains and the zero level curves that define them are of considerable
current interest because they determine many important properties of random
fields [$1$]. \ At present there is little information available on the nodal
domains of ordinary \emph{vector} speckle; this prevents us from using our
present methods to calculate these domains for vector speckled speckle, and
below we consider nodal domains of the phase only of scalar speckled speckle.

\medskip

\emph{Densities of zero crossings and densities of intersections of these
crossings with a straight line.}

\medskip

Phase extrema, saddles, and vortices [$16$], are all located at intersections
of the zero level crossings Z$_{x}$ and Z$_{y}$ of the first order phase
derivatives $\varphi_{x}$ and $\varphi_{y}$. \ It is therefore reasonable to
expect that like the number density of extrema, E, which shows an anomalous
peak in Fig. \ref{Fig3} at $K\sim0.13$, also the line densities $D_{x}%
=D_{y}=D$ of Z$_{x}$ and Z$_{y}$ would show similar anomalous peaks.

For many types of random curves in an isotropic field there is a simple
relationship [$15$] between $\mathcal{D}$ and the density $N$ of intersections
of these curves with a straight line, say the $x$-axis,%
\begin{equation}
\mathcal{D}=\frac{\pi}{2}N. \label{Buffon}%
\end{equation}
This relationship, however, does not hold for level curves of derivatives
because, as discussed below for the phase, differentiation destroys the
required isotropy.

\subsubsection{Densities of intersections of zero crossings of phase
derivatives with the $x$-axis.}

\medskip

\emph{General results.}

\medskip

Previously, we calculated for normal speckle the quantities $N_{x}\left(
\varphi_{x},L\right)  $ ($N_{x}\left(  \varphi_{y},L\right)  $), which are the
density of intersections of crossings at level $L$ of phase derivatives
$\varphi_{x}$ ($\varphi_{y}$) with the $x$-axis [$17$].\ Extending by
inspection these results to the zero crossings Z$_{x}$ and Z$_{y}$ of speckled
speckle, we have%
\begin{equation}
N_{x,x}^{\left(  \text{T}_{a}\text{T}_{b}\right)  }=N_{x}^{\left(
\text{T}_{a}\text{T}_{b}\right)  }\left(  \varphi_{x},0\right)  =\frac
{\sqrt{r_{2}^{\left(  \text{T}_{a}\text{T}_{b}\right)  }}}{\pi}\left[
\eta^{\left(  \text{T}_{a}\text{T}_{b}\right)  }\right]  ^{1/2}, \label{NxPx}%
\end{equation}
where $r_{2}^{\left(  \text{T}_{a}\text{T}_{b}\right)  }$ is given in Appendix
B, $\eta^{\left(  \text{T}_{a}\text{T}_{b}\right)  }$ in Appendix D.

Now, in the isotropic random field of interest here, on average the
length/unit area of Z$_{x}$ and Z$_{y}$ must be the same, therefore, if Eq.
(\ref{Buffon}) were to hold we would have $N_{x}^{\left(  \text{T}_{a}%
\text{T}_{b}\right)  }\left(  \varphi_{x},0\right)  =N_{x}^{\left(
\text{T}_{a}\text{T}_{b}\right)  }\left(  \varphi_{y},0\right)  $; but this is
not the case, and instead [$17$]%
\begin{gather}
N_{x,y}^{\left(  \text{T}_{a}\text{T}_{b}\right)  }=N_{x}^{\left(
\text{T}_{a}\text{T}_{b}\right)  }\left(  \varphi_{y},0\right)  =\frac
{\sqrt{r_{2}^{\left(  \text{T}_{a}\text{T}_{b}\right)  }}}{\pi}\nonumber\\
\times E\left(  \sqrt{\frac{2-\eta^{\left(  \text{T}_{a}\text{T}_{b}\right)
}}{3}}\right)  ,\;\eta^{\left(  \text{T}_{a}\text{T}_{b}\right)  }%
\leq2,\nonumber\\
=\frac{\sqrt{r_{2}^{\left(  \text{T}_{a}\text{T}_{b}\right)  }\left(
\eta^{\left(  \text{T}_{a}\text{T}_{b}\right)  }+1\right)  }}{\sqrt{3}\pi
}\nonumber\\
\times E\left(  \sqrt{\frac{\eta^{\left(  \text{T}_{a}\text{T}_{b}\right)
}-2}{\eta^{\left(  \text{T}_{a}\text{T}_{b}\right)  }+1}}\right)
,\;\eta^{\left(  \text{T}_{a}\text{T}_{b}\right)  }\geq2, \label{NxPy}%
\end{gather}
where $E$ is the complete elliptic integral of the second kind.

\bigskip

\emph{Results for a single source.}

\bigskip

Eqs. (\ref{NxPx}) and (\ref{NxPy}) with single superscripts hold, of course,
for ordinary speckle. \ For a Gaussian (G), disk (D), and ring (R),
$\eta^{\left(  G\right)  }=2$, $\eta^{\left(  D\right)  }=1$, $\eta^{\left(
R\right)  }=1/2$, and
\begin{subequations}
\label{NxyGDR}%
\begin{align}
N_{x,x}^{\left(  G\right)  }/N_{x,y}^{\left(  G\right)  }  &
=0.900,\label{NxyG}\\
N_{x,x}^{\left(  D\right)  }/N_{x,y}^{\left(  D\right)  }  &
=0.699,\label{NxyD}\\
N_{x,x}^{\left(  R\right)  }/N_{x,y}^{\left(  R\right)  }  &  =0.524,
\label{NxyR}%
\end{align}
i.e. $N_{x,x}^{\left(  T\right)  }<N_{x,y}^{\left(  T\right)  }$ in all three cases.

\bigskip

\emph{Anomalous reversals for compound sources.}

\bigskip

From Eqs. (\ref{NxPx}), (\ref{NxPy}), and (\ref{r2p}), it follows that
$b^{-1}N_{x,x}$ and $b^{-1}N_{x,y}$ are functions only of $\rho=a/b$ and $K$.
\ Accordingly, in Fig. \ref{Fig9}(a) we plot $b^{-1}N_{x,x}^{\left(
\text{DD}\right)  }$ and $b^{-1}N_{x,y}^{\left(  \text{DD}\right)  }$ vs. $K$
for two disks (DD) for $\rho=0.1$. \ As can be seen, for very small and very
large $K$ (the regions labelled `b' in the figure) $N_{x,x}^{\left(
\text{DD}\right)  }<N_{x,y}^{\left(  \text{DD}\right)  }$, in agreement with
Eq. (\ref{NxyD}). \ In the intermediate region $10^{-4}<K<1$ (region `c'),
however, this situation reverses, and anomalously $N_{x,x}^{\left(
\text{DD}\right)  }>N_{x,y}^{\left(  \text{DD}\right)  }$. \ This reversal is
a general feature, and occurs also for two Gaussians (GG) and two rings (RR),
both of which yield graphs that are quite similar to Fig. \ref{Fig9}(a).

\vspace{0.1in}

\emph{Explanation of anomalous reversals.}

\vspace{0.1in}

The anomalous $N_{x,x}/N_{x,y}$ ratio, as well as the normal ratio, find their
explanation in the structure of the phase field. \ For normal speckle, and in
regions `b' of Fig. \ref{Fig9}(a) for speckled speckle, vortices outnumber
extrema. \ As discussed in [$16$], in the immediate vicinity of a vortex
Z$_{x}$ (Z$_{y}$) is necessarily strictly parallel to the $x$-axis ($y$-axis),
Fig. \ref{Fig9}(b). \ As a result the number of Z$_{x}$ (Z$_{y}$) that
\emph{cross} the $x$-axis ($y$-axis) is reduced (increased), resulting in
$N_{x,x}<N_{x,y}$. \ But in region `c', extrema outnumber vortices, and
because, as discussed in [$18$], at extrema there is a strong tendency for
Z$_{x}$ (Z$_{y}$) to be biased perpendicular (parallel) to the $x$-axis, Fig.
\ref{Fig9}(c), the number of Z$_{x}$ (Z$_{y}$) that cross the $x$-axis is
increased (reduced), resulting in $N_{x,x}>N_{x,y}$.%

\begin{figure}
[pb]
\begin{center}
\includegraphics[
natheight=7.395000in,
natwidth=5.819300in,
height=2.8643in,
width=2.4137in
]%
{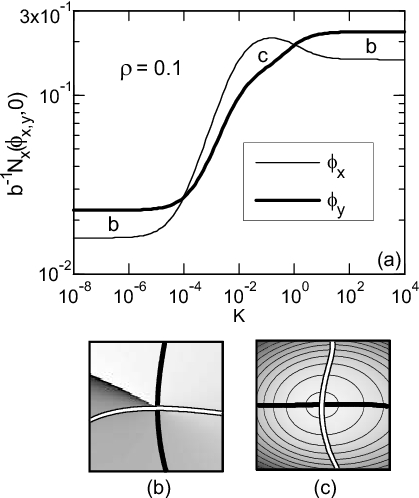}%
\caption{(a) Densities of zero level crossing with the $x$-axis for two disks
(DD). \ Thin curve, $b^{-1}N_{x}^{\left(  \text{DD}\right)  }\left(
\varphi_{x},0\right)  $, Eq. (\ref{NxPx}), thick curve, $b^{-1}N_{x}^{\left(
\text{DD}\right)  }\left(  \varphi_{y},0\right)  $, Eq. (\ref{NxPy}). \ Here
$\rho=a/b=0.1$. \ (b),(c) Zero crossings Z$_{x}$ of $\varphi_{x}$ ( Z$_{y}$ of
$\varphi_{y}$), shown as thick white (black) lines, superimposed on a map of
(b) a phase vortex, (c) a phase maximum. \ In both (b) and (c) the phase
increases $0-2\pi$ black to white. \ For vortices, Z$_{x}$ (Z$_{y}$) is always
strictly parallel (perpendicular) to the $x$-axis. \ For extrema, the
orientation of these zero crossings depends on the anisotropy and orientation
of the elliptical contours in the immediate vicinity of the extremum. \ For
circular contours, and for elliptical contours whose principal axes parallel
the $xy$-axis system, Z$_{x}$ (Z$_{y}$) is parallel to the $y$-axis
($x$-axis). \ For other cases there is a strong bias towards an orientation in
which the absolute value of the angle that Z$_{x}$ (Z$_{y}$) makes with the
$x$-axis is greater than (less than) $45^{o}$. \ }%
\label{Fig9}%
\end{center}
\end{figure}

\bigskip

\emph{Relationship between densities of phase critical points and zero
crossings of phase derivatives.}

\bigskip

Eq. (\ref{Buffon}) does not strictly apply to Z$_{x}$ and Z$_{y}$ separately,
nonetheless, because the differences in $N_{x,x}$ and $N_{x,y}$ are not
severe, and in light of the explanation given above for this difference, we
can reasonably assume that Eq. (\ref{Buffon}) does apply, at least
approximately, to the \emph{average} $x$-axis crossing density
\end{subequations}
\begin{equation}
\left\langle N_{x}^{\left(  \text{T}_{a}\text{T}_{b}\right)  }\right\rangle
\equiv\frac{1}{2}\left(  N_{x}^{\left(  \text{T}_{a}\text{T}_{b}\right)
}\left(  \varphi_{x},0\right)  +N_{x}^{\left(  \text{T}_{a}\text{T}%
_{b}\right)  }\left(  \varphi_{y},0\right)  \right)  . \label{aveNx}%
\end{equation}

\begin{figure}
[pt]
\begin{center}
\includegraphics[
natheight=7.949400in,
natwidth=5.832300in,
height=3.6054in,
width=2.6455in
]%
{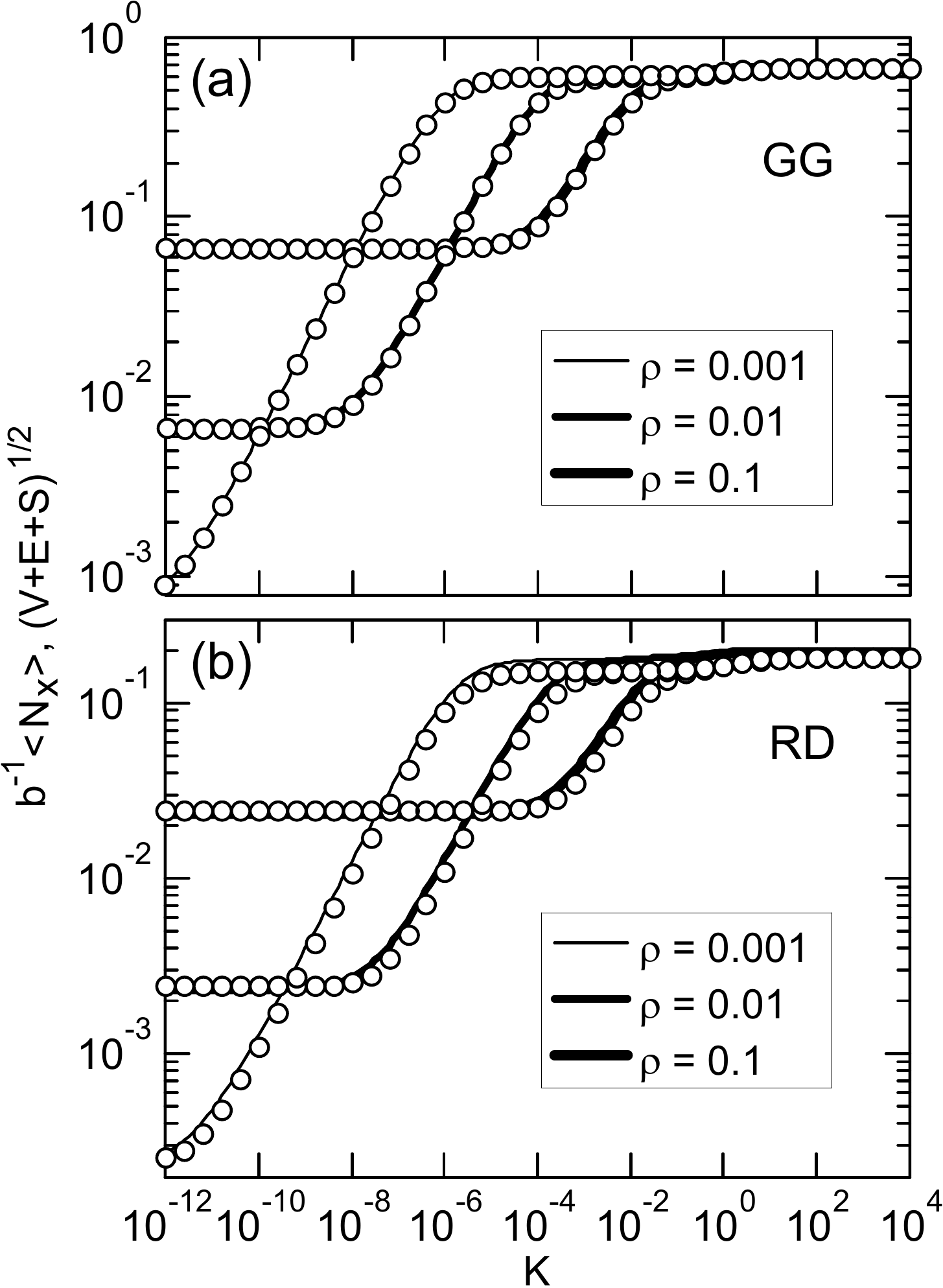}%
\caption{Test of Eq. (\ref{VESNxy}). \ The average zero level crossing
densities, $b^{-1}\left\langle N_{x}\right\rangle $, are shown by continuous
curves, whereas the square roots of the total density of all phase critical
points, $\sqrt{V+S+E}$, are shown by small white circles. \ In (a) the $a$ and
$b$ source fields are both Gaussians (GG). \ In (b) the $a$ source field is a
ring and the $b$ source field is a disk (RD).}%
\label{Fig10}%
\end{center}
\end{figure}

Because all phase critical points - extrema, saddles, and vortices - lie on
intersections of Z$_{x}$ and Z$_{y}$, one can reasonably expect some
relationship to exist between the densities of these zero lines and the
density of critical points. \ We therefore attempt to find a relationship
between the total phase critical point density $V+E+S=2S$, Eqs. (\ref{pEV})
and (\ref{pSEV}), and $\left\langle N_{x}^{\left(  \text{T}_{a}\text{T}%
_{b}\right)  }\right\rangle $. \ Noting that $V$, $S$, and $E$ for normal
speckle are proportional to $p^{2}$, with $p$ the characteristic inverse
length scale of the autocorrelation function, whereas $\left\langle
N_{x}\right\rangle $ is proportional to $p$, we make the simplest possible
ansatz%
\begin{align}
\sqrt{\left(  V+E+S\right)  ^{\left(  \text{T}_{a}\text{T}_{b}\right)  }}  &
\approx\frac{1}{2}\left(  c^{\left(  \text{T}_{a}\right)  }+c^{\left(
\text{T}_{b}\right)  }\right) \nonumber\\
&  \times b^{-1}\left\langle N_{x}^{\left(  \text{T}_{a}\text{T}_{b}\right)
}\right\rangle , \label{VESNxy}%
\end{align}
where $c^{\left(  \text{T}_{a}\right)  }$ and $c^{\left(  \text{T}_{b}\right)
}$ are constants calculated from the single component analog of Eq.
(\ref{VESNxy}) (without the factor of $%
\frac12
)$ for normal $a$ and $b$ speckle fields using the known values of $V$, $E$,
$S$, $N_{x,x}$, and $N_{x,y}$ for these fields. \ In Fig. \ref{Fig10} we plot
the R.H.S. of Eq. (\ref{VESNxy}) as continuous curves on which are
superimposed small circles representing the L.H.S of this equation. $\ $Shown
are representative results for the case of two Gaussians (GG) and the case
where $a$ is a ring and $b$ is a disk (RD). \ Other source combinations that
are not shown yield similar results. \ Considering the wide range of densities
($10^{-4}$ - $1$), the broad parameter range for $K$ ($10^{-12}$ - $10^{4}$),
for $\rho$ ($0.001$ - $0.1$), and for $\eta$ ($0.5$ - $10^{5}$, Appendix D),
and the fact that there are no adjustable parameters, the agreement with Eq.
(\ref{VESNxy}) seems satisfactory.%

\section{CRITICAL POINTS AND ZERO LEVEL CROSSINGS OF THE REAL AND IMAGINARY
PARTS OF THE WAVEFUNCTION}

The real (imaginary) part of the wavefunction of a speckle pattern represents
an archetypal Gaussian random surface. \ The critical points that define the
major features of such a surface are its extrema (maxima plus minima), saddle
points, and umbilic points, the latter being points of isotropic curvature;
these are frequently located on the sloping sides of extrema. \ Here we study
the number densities of these features, and the densities of the associated
level crossings, for speckled speckle. \ In what follows we consider the real
($\mathcal{R}$) part of the wavefunction, but all results hold equally well
for the imaginary ($\mathcal{I}$) part.

\subsection{Critical Points}

\bigskip

\emph{General results.}

\bigskip

The number density $\mathcal{E}$ of extrema [$19$], and the density
$\mathcal{U}$ of umbilics [$20$], in normal speckle is extended by inspection
to speckled speckle as%
\begin{equation}
\mathcal{E}^{\left(  \text{T}_{a}\text{T}_{b}\right)  }=\frac{1}{3\sqrt{3}\pi
}\frac{r_{4}^{\left(  \text{T}_{a}\text{T}_{b}\right)  }}{r_{2}^{\left(
\text{T}_{a}\text{T}_{b}\right)  }}, \label{Ereal}%
\end{equation}%
\begin{equation}
\mathcal{U}^{\left(  \text{T}_{a}\text{T}_{b}\right)  }=\frac{1}{4\pi}%
\frac{M_{6}^{\left(  \text{T}_{a}\text{T}_{b}\right)  }}{M_{4}^{\left(
\text{T}_{a}\text{T}_{b}\right)  }}. \label{Ureal}%
\end{equation}

\bigskip

\emph{Results for single sources.}

\bigskip

For normal speckle [$19,20$] Eqs. (\ref{Ereal}) and (\ref{Ureal}) with $k=0$,
yield
\begin{subequations}
\begin{align}
\mathcal{U}^{\left(  \text{G}\right)  }  &  =\frac{3}{\pi}p^{2},\;\mathcal{E}%
^{\left(  \text{G}\right)  }=\frac{2\sqrt{3}}{3\pi}p^{2},\label{UEG}\\
\mathcal{U}^{\left(  \text{D}\right)  }  &  =\frac{3}{16\pi}p^{2}%
,\;\mathcal{E}^{\left(  \text{D}\right)  }=\frac{\sqrt{3}}{18\pi}%
p^{2},\label{UED}\\
\mathcal{U}^{\left(  \text{D}\right)  }  &  =\frac{1}{4\pi}p^{2}%
,\;\mathcal{E}^{\left(  \text{D}\right)  }=\frac{1}{4\sqrt{3}\pi}p^{2}.
\label{UER}%
\end{align}
where $1/p$ is the characteristic length scale in the speckle pattern, Eqs.
(\ref{Wpr}). The number density ratios $\mathcal{U}/\mathcal{E}$ for a
Gaussian (G), disk (D), and ring (R), are $2.6$, $1.9$, and $1.7$,
respectively. \ Thus, in normal speckle each extremum (maximum or minimum) is
decorated by approximately one umbilic point; for speckled speckle, however,
the number of umbilic points per extremum can be arbitrarily large, as we now show.

\bigskip

\emph{Anomalous ratio of umbilic points to extrema in speckled speckle.}

\bigskip

Using Eqs. (\ref{Ereal}) and (\ref{Ureal}) we find that for any given small
value of $\rho=a/b$, when $K=\mathcal{K}_{\max}^{\left(  \text{T}_{a}%
\text{T}_{b}\right)  }$, $\left(  \mathcal{U}/\mathcal{E}\right)  _{\max
}^{\left(  \text{T}_{a}\text{T}_{b}\right)  }=\left(  \mathcal{U}^{\left(
\text{T}_{a}\text{T}_{b}\right)  }/\mathcal{E}^{\left(  \text{T}_{a}%
\text{T}_{b}\right)  }\right)  _{\max}$ attains a large, maximum value, where
to leading order,%

\end{subequations}
\begin{subequations}
\label{maxKUE}%
\begin{align}
\mathcal{K}_{\max}^{\left(  \text{T}_{a}\text{T}_{b}\right)  }  &
=\frac{c_{4}^{\left(  \text{T}_{a}\right)  }}{c_{2}^{\left(  \text{T}%
_{a}\right)  }}\rho^{4},\label{UEKmax}\\
\left(  \mathcal{U}/\mathcal{E}\right)  _{\max}^{\left(  \text{T}_{a}%
\text{T}_{b}\right)  }  &  =\frac{9\sqrt{3}}{40}\left(  \frac{c_{2}^{\left(
\text{T}_{a}\right)  }c_{6}^{\left(  \text{T}_{a}\right)  }}{c_{4}^{\left(
\text{T}_{a}\right)  }c_{4}^{\left(  \text{T}_{b}\right)  }}\right)  \frac
{1}{\rho^{2}}, \label{UEmax}%
\end{align}
with $c_{2n}$, $n=1,2,3$, given in Eqs. (\ref{c2n}). \ As an example, for $a$
a ring and $b$ a Gaussian, $\mathcal{K}_{\max}^{\left(  \text{RG}\right)
}=\frac{3}{16}\rho^{4}$, and $\left(  \mathcal{U}/\mathcal{E}\right)  _{\max
}^{\left(  \text{RG}\right)  }=3\sqrt{3}/\rho^{2}$.

For T = G, D, R, $\left(  \mathcal{U}/\mathcal{E}\right)  _{\max}^{\left(
\text{TT}\right)  }/\left(  \mathcal{U}/\mathcal{E}\right)  ^{\left(
\text{T}\right)  }=1/\left(  4\rho^{2}\right)  $. For $\rho=0.1$, for example,
the ratio of umbilic points to extrema is therefore enhanced by a factor of
$25$.

In Fig. \ref{Fig11}(a) we plot $\mathcal{U}$, $\mathcal{E}$, and
$\mathcal{U}/\mathcal{E}$ for two Gaussians (GG) for $\rho=0.1$. \ As can be
seen, both $\mathcal{U}$ and $\mathcal{E}$ extrapolate smoothly between their
limiting $a$ and $b$ field values, whereas $\mathcal{U}/\mathcal{E}$ shows the
expected large peak at $K=10^{-4}$. \ These curves are representative, and
similar results are obtained for other GDR combinations. \ In Figs.
\ref{Fig11}(b) and (c) we show the real part $\mathcal{R}$ of the wavefield
with its umbilic points superimposed. \ These points were located using the
fact that they lie at the intersections of the zero [$20$] lines of the
Stokes-like parameters $\Sigma_{1}$ and $\Sigma_{2}$, where
\end{subequations}
\begin{subequations}
\label{s012}%
\begin{align}
\Sigma_{1}  &  =\mathcal{R}_{xx}-\mathcal{R}_{yy},\label{s1}\\
\Sigma_{2}  &  =2\mathcal{R}_{xy}, \label{s2}%
\end{align}
and the subscripts imply differentiation with respect to $x$ and $y$.
\begin{figure}
[ptb]
\begin{center}
\includegraphics[
natheight=4.337900in,
natwidth=5.892800in,
height=3.2102in,
width=2.6775in
]%
{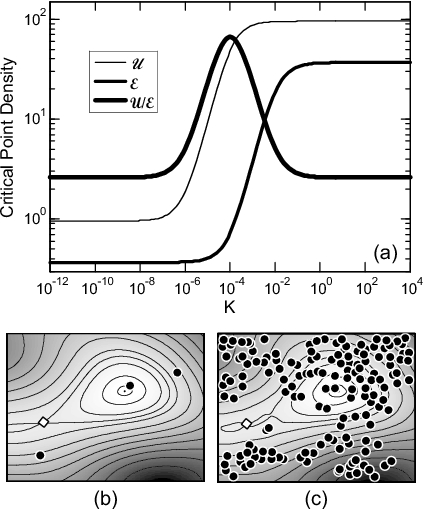}%
\caption{Extrema and umbilic points for two Gaussians (GG) for $\rho=a/b=0.1$.
\ (a) Densities $\mathcal{U}$ of umbilic points, $\mathcal{E}$ of extrema, and
their ratio $\mathcal{U}/\mathcal{E}$. \ (b),(c) Gray scale coded contour maps
of the real part of the wavefunction $\mathcal{R}$ with superimposed umbilic
points (small black circles). \ $\mathcal{R}$ increases black to white; the
small white diamond marks a saddle point. \ (b) $I_{b}=0$ ($K=0$). \ \ (c)
$I_{b}/I_{a}=10^{-6}$ ($K=10^{-4}$). \ Note that the underlying stationary
point structure (extrema and saddle point) in (b) and (c) are very nearly the
same. \ This is typical, and even for $K\sim1$ the underlying stationary point
structure of the $a$ field remains visible \ The $152$ umbilic points in (c),
which corresponds to the peak of $\mathcal{U}/\mathcal{E}$ in (a), are
consistent with the number expected for a region that contains a central
extremum (here a maximum), and parts of the sloping sides of two other extrema
(here minima located in the upper left and lower right corners of the figure).
\ The hyperbolic curvature of the saddle point tends to inhibit formation of
umbilic points because the principal radii of curvature of the saddle have
opposite signs, whereas at an umbilic point both radii must be
equal.\ \ \ \ \ }%
\label{Fig11}%
\end{center}
\end{figure}

\subsection{Zero Level Crossings}

\emph{General results.}

Level crossing densities for $\mathcal{R}$, and its first, $\mathcal{R}%
_{x},\mathcal{R}_{y}$, and second, $\mathcal{R}_{xx},\mathcal{R}%
_{xy},\mathcal{R}_{yy}$, derivatives have long been known for normal speckle
[$17,21$], and we obtain by inspection for level zero for speckled speckle,
\end{subequations}
\begin{subequations}
\label{NxRRxRxx}%
\begin{align}
\mathcal{N}_{0}^{\left(  \text{T}_{a}\text{T}_{b}\right)  }  &  \equiv
\mathcal{N}_{x}^{\left(  \text{T}_{a}\text{T}_{b}\right)  }\left(
\mathcal{R},0\right)  =\frac{1}{\pi}\sqrt{r_{2}^{\left(  \text{T}_{a}%
\text{T}_{b}\right)  }},\label{NxR}\\
\mathcal{N}_{1}^{\left(  \text{T}_{a}\text{T}_{b}\right)  }  &  \equiv
\mathcal{N}_{x}^{\left(  \text{T}_{a}\text{T}_{b}\right)  }\left(
\mathcal{R}_{x},0\right)  =\frac{1}{\pi}\sqrt{r_{4}^{\left(  \text{T}%
_{a}\text{T}_{b}\right)  }/r_{2}^{\left(  \text{T}_{a}\text{T}_{b}\right)  }%
},\label{NxRx}\\
\mathcal{N}_{2}^{\left(  \text{T}_{a}\text{T}_{b}\right)  }  &  \equiv
\mathcal{N}_{x}^{\left(  \text{T}_{a}\text{T}_{b}\right)  }\left(
\mathcal{R}_{xx},0\right)  =\frac{1}{\pi}\sqrt{r_{6}^{\left(  \text{T}%
_{a}\text{T}_{b}\right)  }/r_{4}^{\left(  \text{T}_{a}\text{T}_{b}\right)  }},
\label{NxRxx}%
\end{align}

and
\end{subequations}
\begin{subequations}
\label{NxRyRxyRyy}%
\begin{align}
\mathcal{N}_{x}^{\left(  \text{T}_{a}\text{T}_{b}\right)  }\left(
\mathcal{R}_{y},0\right)   &  =\frac{1}{\sqrt{3}}\mathcal{N}_{x}^{\left(
\text{T}_{a}\text{T}_{b}\right)  }\left(  \mathcal{R}_{x},0\right)
,\label{NxRy}\\
\mathcal{N}_{x}^{\left(  \text{T}_{a}\text{T}_{b}\right)  }\left(
\mathcal{R}_{xy},0\right)   &  =\sqrt{\frac{3}{5}}\mathcal{N}_{x}^{\left(
\text{T}_{a}\text{T}_{b}\right)  }\left(  \mathcal{R}_{xx},0\right)
,\label{NxRxy}\\
\mathcal{N}_{x}^{\left(  \text{T}_{a}\text{T}_{b}\right)  }\left(
\mathcal{R}_{yy},0\right)   &  =\sqrt{\frac{1}{5}}\mathcal{N}_{x}^{\left(
\text{T}_{a}\text{T}_{b}\right)  }\left(  \mathcal{R}_{xx},0\right)  .
\label{NxRyy}%
\end{align}

For the ratios of zero crossing densities we have%
\end{subequations}
\begin{equation}
\mathcal{N}_{1,0}^{\left(  \text{T}_{a}\text{T}_{b}\right)  }\equiv
\frac{\mathcal{N}_{1}^{\left(  \text{T}_{a}\text{T}_{b}\right)  }}%
{\mathcal{N}_{0}^{\left(  \text{T}_{a}\text{T}_{b}\right)  }}=\sqrt
{\eta^{\left(  \text{T}_{a}\text{T}_{b}\right)  }+1}, \label{N1,0}%
\end{equation}
where $\eta^{\left(  \text{T}_{a}\text{T}_{b}\right)  }$ is given in Appendix
D, and%
\begin{equation}
\mathcal{N}_{2,1}^{\left(  \text{T}_{a}\text{T}_{b}\right)  }\equiv
\frac{\mathcal{N}_{2}^{\left(  \text{T}_{a}\text{T}_{b}\right)  }}%
{\mathcal{N}_{1}^{\left(  \text{T}_{a}\text{T}_{b}\right)  }}=\frac{\sqrt{10}%
}{3^{3/4}}\sqrt{\frac{\mathcal{U}^{\left(  \text{T}_{a}\text{T}_{b}\right)  }%
}{\mathcal{E}^{\left(  \text{T}_{a}\text{T}_{b}\right)  }}}. \label{N2,1}%
\end{equation}

\bigskip

\emph{Results for single sources.}

\bigskip

In normal speckle fields zero crossings of $R$, $R_{x}$, and $R_{xx}$ have
similar densities. \ Setting $K=0$ ($I_{b}=0$) in Eqs. (\ref{N1,0}) and
(\ref{N2,1}), we have in obvious notation
\begin{subequations}
\label{N12}%
\begin{align}
\mathcal{N}_{1,0}^{\left(  \text{G}\right)  }  &  =\sqrt{3}\approx
1.73,\;\mathcal{N}_{2,1}^{\left(  \text{G}\right)  }=\sqrt{5/3}\approx
1.29,\label{N12G}\\
\mathcal{N}_{1,0}^{\left(  \text{D}\right)  }  &  =\sqrt{2}\approx
1.41,\;\mathcal{N}_{2,1}^{\left(  \text{G}\right)  }=\sqrt{5}/2\approx
1.12,\label{N12D}\\
\mathcal{N}_{1,0}^{\left(  \text{R}\right)  }  &  =\sqrt{3/2}\approx
1.22,\;\mathcal{N}_{2,1}^{\left(  \text{R}\right)  }=\sqrt{10}/3\approx1.05,
\label{N12R}%
\end{align}
For speckled speckle, however, these ratios can be arbitrarily large, as we
now show.

\bigskip

\emph{Anomalous ratios for compound sources.}

$\medskip$

$\mathcal{N}_{1,0}^{\left(  \text{T}_{a}\text{T}_{b}\right)  }$, Eq.
(\ref{N1,0}), and $\mathcal{N}_{2,1}^{\left(  \text{T}_{a}\text{T}_{b}\right)
}$, Eq. (\ref{N2,1}), can be made arbitrarily large by a suitable choice of
$K^{\left(  \text{T}_{a}\text{T}_{b}\right)  }$ and $\rho$. \ For
$\mathcal{N}_{1,0}^{\left(  \text{T}_{a}\text{T}_{b}\right)  }$ we find that
for any given small value of $\rho$, when $K^{\left(  \text{T}_{a}\text{T}%
_{b}\right)  }=\left(  \mathcal{K}_{1,0}^{\left(  \text{T}_{a}\text{T}%
_{b}\right)  }\right)  _{\max}$, $\mathcal{N}_{1,0}^{\left(  \text{T}%
_{a}\text{T}_{b}\right)  }$ attains a maximum value, $\left(  \mathcal{N}%
_{1,0}^{\left(  \text{T}_{a}\text{T}_{b}\right)  }\right)  _{\max}$, where to
leading order in $\rho$,
\end{subequations}
\begin{subequations}
\label{KN10max}%
\begin{align}
\left(  \mathcal{K}_{1,0}^{\left(  \text{T}_{a}\text{T}_{b}\right)  }\right)
_{\max}  &  =\frac{c_{2}^{\left(  \text{T}_{a}\right)  }}{c_{2}^{\left(
\text{T}_{b}\right)  }}\rho^{2},\label{K10max}\\
\left(  \mathcal{N}_{1,0}^{\left(  \text{T}_{a}\text{T}_{b}\right)  }\right)
_{\max}  &  =\frac{1}{2\rho}\sqrt{\frac{c_{4}^{\left(  \text{T}_{b}\right)  }%
}{c_{2}^{\left(  \text{T}_{a}\right)  }c_{2}^{\left(  \text{T}_{b}\right)  }}%
}. \label{N10max}%
\end{align}
Similarly, for $\mathcal{N}_{2,1}^{\left(  \text{T}_{a}\text{T}_{b}\right)  }$
we find
\end{subequations}
\begin{subequations}
\label{KN21max}%
\begin{align}
\left(  \mathcal{K}_{2,1}^{\left(  \text{T}_{a}\text{T}_{b}\right)  }\right)
_{\max}  &  =\frac{c_{4}^{\left(  \text{T}_{a}\right)  }}{c_{4}^{\left(
\text{T}_{b}\right)  }}\rho^{4},\label{K21max}\\
\left(  \mathcal{N}_{2,1}^{\left(  \text{T}_{a}\text{T}_{b}\right)  }\right)
_{\max}  &  =\frac{1}{2\rho}\sqrt{\frac{c_{2}^{\left(  \text{T}_{a}\right)
}c_{6}^{\left(  \text{T}_{b}\right)  }}{c_{4}^{\left(  \text{T}_{a}\right)
}c_{4}^{\left(  \text{T}_{b}\right)  }}}. \label{N21max}%
\end{align}
As representative examples we plot in Fig. \ref{Fig12} $\mathcal{N}%
_{1,0}^{\left(  \text{GG}\right)  }$ and $\mathcal{N}_{2,1}^{\left(
\text{GG}\right)  }$ for $\rho=0.01$. \ As can be seen, both densities show
large enhancements at their peaks over their normal speckle values. \ Below we
inquire as to the sources of these enhancements.%

\begin{figure}
[pb]
\begin{center}
\includegraphics[
natheight=4.337900in,
natwidth=5.849600in,
height=1.9804in,
width=2.6593in
]%
{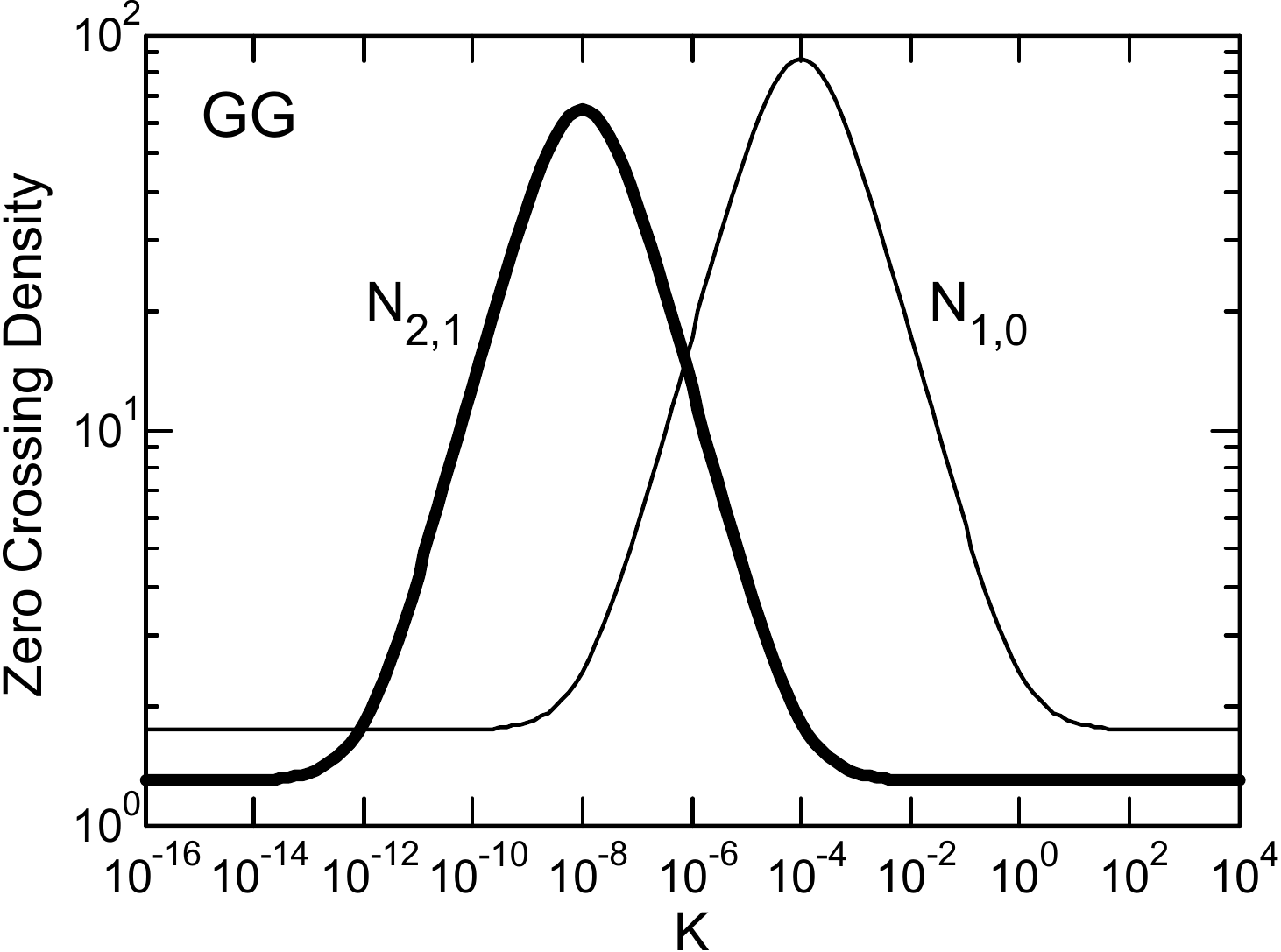}%
\caption{Zero crossing densities $N_{1,0}^{\left(  GG\right)  }$, Eq.
(\ref{N10max}), and $N_{2,1}^{\left(  GG\right)  }$, Eq. (\ref{N21max}), for
two Gaussians (GG) for $\rho=0.01$. \ For this value of $\rho$, $N_{1,0}%
^{\left(  GG\right)  }$ ($N_{2,1}^{\left(  GG\right)  }$) is enhanced at its
peak, $K=10^{-4}$ ($K=10^{-8}$), by a factor of $50$ ($44$) over its normal
speckle value, Eqs. (\ref{N12G}). }%
\label{Fig12}%
\end{center}
\end{figure}

$\bigskip$

$\mathcal{N}_{1,0}^{\left(  \text{T}_{a}\text{T}_{b}\right)  }$.

\bigskip

The zero crossings of $\mathcal{R}$, which obey Eq. (\ref{Buffon}), divide the
field into nodal domains of opposite sign. \ In normal speckle maxima (minima)
mostly occupy positive (negative) domains. $\mathcal{N}_{x}\left(
\mathcal{R},0\right)  $ therefore roughly scales with the square root of the
density of extrema. \ The same is true for $\mathcal{N}_{x}\left(
\mathcal{R}_{x},0\right)  $, because extrema (and saddles) lie at
intersections of $\mathcal{R}_{x}$ and $\mathcal{R}_{y}$. \ Accordingly, in
normal speckle $\mathcal{N}_{x}\left(  \mathcal{R}_{x},0\right)
\sim\mathcal{N}_{x}\left(  \mathcal{R},0\right)  $, Eqs. (\ref{N12}). \ In
speckled speckle, however, the situation is very different.

For vanishingly small $K$ the only extrema are those of the $a$ field, and the
nodal domains of $\mathcal{R}$ are determined by these extrema. As $K$
increases, however, numerous very small amplitude extrema associated with the
presence of the $b$ field (here colloquially \textquotedblleft$b$
extrema\textquotedblright) begin to appear. \ These $b$ extrema decorate with
equal probability both positive and negative nodal domains of $\mathcal{R}$,
but because of their very small amplitudes they do not divide up these nodal
domains. \ Thus, $\mathcal{N}_{0}=\mathcal{N}_{x}\left(  \mathcal{R},0\right)
$ is not significantly increased by the presence of the $b$ extrema.
\ $\mathcal{N}_{1}=\mathcal{N}_{x}\left(  \mathcal{R}_{x},0\right)  $,
however, is increased in proportion to the square root of the density of these
extrema, because $\mathcal{R}_{x}$ (and $\mathcal{R}_{y}$) goes to zero at an
extremum no matter how small its amplitude. \ As a result, $\mathcal{N}%
_{1,0}=$ $\mathcal{N}_{1}/\mathcal{N}_{0}$ begins to grow with $K$. \ This
growth continues until the amplitude of the $b$ source field is such that the
amplitudes of the $b$ extrema become large enough to carve up the nodal
domains of $\mathcal{R}$ into small regions. \ When this happens
$\mathcal{N}_{0}$ begins to grow rapidly. \ Thus the peak in $\mathcal{N}%
_{1,0}$ is due to an initial growth of $\mathcal{N}_{1}$ with no significant
growth in $\mathcal{N}_{0}$, followed by a leveling off of $\mathcal{N}_{1}$
together with rapid growth in $\mathcal{N}_{0}$. \ This scenario, which
applies to all GDR source combinations, is illustrated in Fig. \ref{Fig13}.

$\bigskip$

$\mathcal{N}_{2,1}^{\left(  \text{T}_{a}\text{T}_{b}\right)  }$.

\bigskip

Why is $\mathcal{N}_{2}=\mathcal{N}_{x}\left(  \mathcal{R}_{xx},0\right)  $
for speckled speckle so much larger than $\mathcal{N}_{1}=\mathcal{N}%
_{x}\left(  \mathcal{R}_{x},0\right)  $, when for normal speckle the two are
nearly equal, Eqs. (\ref{N12})? \ The answer is analogous to the reason,
discussed above, why in speckled speckle $\mathcal{N}_{1}$ is so much larger
than $\mathcal{N}_{0}$, even though in normal speckle also these two densities
are nearly equal.

As discussed at the beginning of this section, for very small $K$ each
extremum is decorated on average by a single umbilic point, but as $K$
increases the number of umbilics/extremum increases rapidly, Eqs.
(\ref{maxKUE}) and Fig. \ref{Fig11}. \ Now, at an umbilic point $\mathcal{R}%
_{xx}=\mathcal{R}_{yy}$ and $\mathcal{R}_{xy}=0$ [$20$]. \ Accordingly, as the
number of umbilics increases so does $\mathcal{N}_{x}\left(  \mathcal{R}%
_{xy},0\right)  $, the number of zero crossings of $\mathcal{R}_{xy}$. \ But
for the circular Gaussian statistics relevant here $\mathcal{N}_{2}=\sqrt
{5/3}\mathcal{N}_{x}\left(  \mathcal{R}_{xy},0\right)  $, Eq. (\ref{NxRxy}),
so that as the number of umbilics increases so also does $\mathcal{N}_{2}$.
\ The net result is that because $\mathcal{N}_{2}$ increases rapidly with the
number of umbilics, whereas $\mathcal{N}_{1}$ increases more slowly, only with
the number of extrema (Fig. \ref{Fig13}), $\mathcal{N}_{2}/\mathcal{N}_{1}$
peaks at some intermediate value of $K$. \ These considerations could be
illustrated by the analog of Fig. \ref{Fig13} in which $\mathcal{N}_{0}$\ is
replaced by $\mathcal{N}_{1}$, $\mathcal{N}_{1}$\ by $\mathcal{N}_{2}$, and
$\mathcal{N}_{1,0}$\ by $\mathcal{N}_{2,1}$; we refrain from presenting the
resulting figure, however, because except for scaling factors it is very
similar to Fig. \ref{Fig13}.%

\begin{figure}
[pt]
\begin{center}
\includegraphics[
natheight=7.799700in,
natwidth=5.903200in,
height=3.7715in,
width=2.8608in
]%
{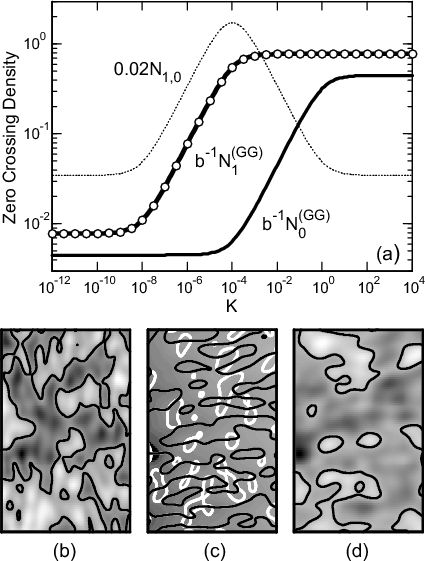}%
\caption{(a) Zero crossings densities $b^{-1}N_{1}$ (thick curve) and
$b^{-1}N_{0}$ (thin curve) vs. $K$ for $a$ and $b$ fields Gaussians (GG) with
$\rho=a/b=0.01$. \ $N_{0}$ ($N_{1}$) is given in Eq. (\ref{NxR}), (Eq.
(\ref{NxRx})). \ The small white circles are the square root of the scaled
density $\mathcal{E}$ of extrema calculated from the exact result
$\sqrt{\mathcal{E}\left(  K\right)  }=\left(  \sqrt{\pi}/3^{3/4}\right)
N_{1}\left(  K\right)  $, where $\mathcal{E}$ is given in Eq. (\ref{Ereal}).
\ \ The dotted curve is $N_{1,0}=N_{1}/N_{0}$ rescaled to permit easy
comparison with curves $b^{-1}N_{0}$ and $b^{-1}N_{1}$. \ (b)-(d) Gray scale
coded maps of $\mathcal{R}\left(  x,y\right)  $ for $\rho=0.01$ for different
values of $K$. \ In all maps $\mathcal{R}$ increases black to white. (b)
$K=10^{-4}$. \ The thick curves are zero crossings of $\mathcal{R}$ that
divide the field into positive/negative nodal domains. \ As can be seen,
maxima (minima) almost invariably lie in positive (negative) domains. \ Maps
for $K<10^{-4}$ differ negligibly from the one shown here. \ (c) The central
region in (b) enlarged by a factor of $100$, so that the area shown is
$10^{-4}$ the area in (b). \ Thick white (black) curves are zero crossings
of\ derivatives $\mathcal{R}_{x}$ ($\mathcal{R}_{y}$). Every intersection of
$\mathcal{R}_{x}$ and $\mathcal{R}_{y}$ marks a stationary point - extremum or
saddle point; on average there are equal numbers of each type. \ Here the
amplitudes of the extrema are so small that they are not apparent visually.
\ (d) The region in (c) for $K=1$. \ Thick curves are zero crossings of
$\mathcal{R}$. \ Here the extrema have large amplitudes. \ Because of this
they are not only visible, but they also divide each nodal domain in (b) into
some $10^{4}$ small domains. \ Maps for $K>1$ differ unimportantly from the
one shown here. \ }%
\label{Fig13}%
\end{center}
\end{figure}

\section{SUMMARY}

A priori, one might reasonably expect that adding a weak random field, the $b$
field, onto a strong random field, the $a$ field, would result in only minor
perturbations of the $a$ field structure and statistics. \ The opposite was
found to be true, however, and just about every structural and statistical
property of the combined field was shown to display large anomalies when the
optical power $P_{b}$ of the $b$ field is some $10^{-4}$ - $10^{-6}$ the power
$P_{a}$ of the $a$ field; these anomalies of speckled speckle are summarized
below for $\rho=a/b=10^{-1}$ - $10^{-2}$.

(\emph{i}) \emph{Singularity clustering}: \ Vortices in scalar fields and C
points in vector fields cluster in the dark regions surrounding the bright
speckle spots of the $a$ field; the reason is that singularity formation
requires equal amplitudes for the $a$ and $b$ fields.

(\emph{ii}) \emph{Phase field statistics}: \ In both scalar and vector fields
extrema can outnumber singularities by orders of magnitude; in contrast, in
normal speckle singularities always outnumber extrema.

\hspace{0.2in}(\emph{a}) \emph{Scalar fields}.\ \ For a sufficiently weak $b$
field, say $P_{b}/P_{a}\sim10^{-8}$ - $10^{-10}$, extrema and vortices
maintain their normal ratio. \ But as $P_{b}$ grows, the number density of
extrema, $E$, grows faster than the density, $V$, of vortices, and the ratio
$E/V$ can become orders of magnitude larger than normal. \ When $P_{b}%
/P_{a}\gg1$, the $E/V$ ratio returns to its normal value. \ Thus, $E/V$ is
normal for both very small and very large $P_{b}/P_{a}$, and peaks at some
still small, intermediate value of $P_{b}/P_{a}$.

\hspace{0.2in}(\emph{b}) \emph{Vector fields}.\ \ For all nonzero $P_{b}$ less
than some critical value $P_{b}^{0}$, the ratio of the number density of
azimuthal extrema, $\mathbb{E}$, can exceed the number density, $\mathbb{V}$,
of C points (Stokes vortices) by orders of magnitude. \ For $P_{b}<P_{b}^{0}$,
$\mathbb{E}/\mathbb{V}$ is essentially independent of $P_{b}$ and is
substantially equal to half the \emph{maximum} scalar field ratio $E/V$ for
the same value of $\rho$;\ $P_{b}^{0}$ approximately equals $P_{b}$ at the
peak of $E/V$. \ When $P_{b}/P_{a}\gg1$, $\mathbb{E}/\mathbb{V}$ returns to
its normal value. \ Thus, $\mathbb{E}/\mathbb{V}$ is large for small $P_{b}$
and small for large $P_{b}$.

(\emph{iii})\emph{ Phase field derivative zero crossings:}\ The ratio of the
density of $x$-derivative zero crossings to $y$-derivative zero crossings
changes anomalously with $P_{b}$; this anomaly is explained by the anomalous
$E/V$ ratio and the differences in the geometry of the zero crossings at
vortices and at extrema.

(\emph{iv})\emph{ Real and imaginary parts of the wave function:}

\hspace{0.2in}(\emph{a}) \emph{Umbilic points. \ }For a sufficiently weak $b$
field, say $P_{b}/P_{a}\sim10^{-10}$ - $10^{-12}$, extrema and umbilic points
maintain their normal $\sim1:1$ ratio. \ But as $P_{b}$ is increased, the
number density of extrema, $\mathcal{E}$, grows faster than the density,
$\mathcal{U}$, of umbilic points, and the ratio $\mathcal{E}/\mathcal{U}$ can
become orders of magnitude larger than normal. \ $\mathcal{E}/\mathcal{U}$
returns to its normal value when $P_{b}/P_{a}\gg1$.

\hspace{0.2in}(\emph{b})\emph{ Nodal domains and derivative zero crossings.
}The densities of the zero crossings that define nodal domains and the
densities of the zero crossings of all first and second order derivatives show
anomalously large ratios. \ These anomalous ratios are explained by the
anomalous $\mathcal{E}/\mathcal{U}$ ratio and the relationships of the
different zero crossings to extrema and umbilic points.

In general, the anomalous ratios of critical point densities occur because the
number density of one wavefield feature grows rapidly with an increase in the
power of the $b$ field while the growth rate of a second feature lags behind.
\ Such an imbalance in growth rates can be expected to be the norm, and other
properties of speckled speckle, including properties that defy calculation,
can be expected to show large anomalies; these anomalies can, and should be,
sought experimentally.
\end{subequations}
\begin{acknowledgments}
D. A. Kessler acknowledges the support of the Israel Science Foundation.
\end{acknowledgments}

\appendix

\section{Composite Field Autocorrelation Function}

Returning to the derivation of the autocorrelation functions of the simple
sources in Eqs. (\ref{SGauss})-(\ref{SDelta}) we write $W$ in terms of the
numerator $N\left(  pr\right)  $ and denominator $D\left(  p\right)  $ of Eq.
(\ref{VCZ}) as%
\begin{equation}
W\left(  pr\right)  =N\left(  pr\right)  /D\left(  p\right)  , \label{NP}%
\end{equation}
where for the Gaussian (G), disk (D), and ring (R):
\begin{subequations}
\begin{align}
N^{\left(  \text{G}\right)  }\left(  pr\right)   &  =2p^{2}exp\left(
-p^{2}r^{2}\right)  ,\;D^{\left(  \text{G}\right)  }\left(  p\right)
=2p^{2};\label{NPG}\\
N^{\left(  \text{D}\right)  }\left(  pr\right)   &  =p^{2}J_{1}\left(
pr\right)  /\left(  pr\right)  ,\;D^{\left(  \text{D}\right)  }\left(
p\right)  =p^{2}/2;\label{NPD}\\
N^{\left(  \text{R}\right)  }\left(  pr\right)   &  =\varepsilon pJ_{0}\left(
pr\right)  ,\;D^{\left(  \text{R}\right)  }\left(  p\right)  =\varepsilon p.
\label{NPR}%
\end{align}
Inserting Eqs. (\ref{NPG})-(\ref{NPR}) into Eq. (\ref{NP}) immediately leads
to Eqs. (\ref{WG})-(\ref{WR}).

The numerator (denominator) of the composite autocorrelation function in Eq.
(\ref{VCZ}) is a linear combination of the $N$ ($D$) in Eqs. (\ref{NPG}%
)-(\ref{NPR}) with $p$ replaced by $a$ or $b$, as required. Inserting the
appropriate combinations of $N$ and $D$ into Eq. (\ref{VCZ}) leads to Eq.
(\ref{Wab}) with%
\end{subequations}
\begin{equation}
K^{\left(  \text{T}_{a}\text{T}_{b}\right)  }=P_{b}^{\left(  \text{T}%
_{b}\right)  }/P_{a}^{\left(  \text{T}_{a}\right)  }, \label{KPbPa}%
\end{equation}
with $P_{a}^{\left(  \text{T}_{a}\right)  }$ ($P_{b}^{\left(  \text{T}%
_{b}\right)  }$) the total optical power in source $a$ ($b$). \ For a Gaussian
(G), disk (D), and ring (R),
\begin{subequations}
\label{PGDR}%
\begin{align}
P_{p}^{\left(  \text{G}\right)  }  &  =4\pi p^{2}I_{p},\label{PG}\\
P_{p}^{\left(  \text{D}\right)  }  &  =\pi p^{2}I_{p},\label{PD}\\
P_{p}^{\left(  \text{R}\right)  }  &  =2\pi p\varepsilon I_{p}, \label{PR}%
\end{align}
where $\ p=a,b.$ \ For a Gaussian $I_{p}$ is the peak intensity at the center
of the source profile, for a disk $I_{p}$ is the uniform source intensity
within the disk, and for a ring $I_{p}$ is the uniform source intensity within
the narrow annulus.

As a specific example, for a composite source with $a$ a Gaussian and $b$ a
ring,%
\end{subequations}
\begin{align*}
W^{\left(  \text{GR}\right)  }\left(  r\right)   &  =\dfrac{\exp\left(
-a^{2}r^{2}\right)  +K^{\left(  \text{GR}\right)  }J_{0}\left(  br\right)
}{1+K^{\left(  \text{GR}\right)  }},\\
K^{\left(  \text{GR}\right)  }  &  =b\varepsilon I_{b}/2a^{2}I_{a}.
\end{align*}

\section{Derivatives}

The autocorrelation function $\mathcal{W}$ of $\mathcal{R}$ (or equivalently
of $\mathcal{I}$) is
\begin{subequations}
\label{rhoOFr}%
\begin{align}
\mathcal{W}\left(  pr\right)   &  =\left\langle \mathcal{R}\left(  0\right)
\mathcal{R}\left(  pr\right)  \right\rangle /\left\langle \mathcal{R}%
^{2}\right\rangle ,\\
&  =\left\langle \mathcal{I}\left(  0\right)  \mathcal{I}\left(  pr\right)
\right\rangle /\left\langle \mathcal{I}^{2}\right\rangle ,\\
&  =\left\langle \mathcal{R}^{2}\right\rangle W\left(  pr\right)
=\left\langle \mathcal{I}^{2}\right\rangle W\left(  pr\right)  .
\end{align}
Suppressing momentarily for notational simplicity superscripts T,
\end{subequations}
\begin{subequations}
\begin{align}
r_{2n}  &  \equiv\left\langle \mathcal{R}_{n}^{2}\right\rangle /\left\langle
\mathcal{R}^{2}\right\rangle ,\\
\left\langle \mathcal{R}_{n}^{2}\right\rangle  &  =\left(  -1\right)
^{n}\left(  \frac{d^{2n}\mathcal{W}\left(  pr\right)  }{dr^{2n}}\right)
_{r=0}. \label{r2n}%
\end{align}
Reintroducing superscripts T,%
\end{subequations}
\begin{equation}
r_{2n}^{\left(  \text{T}_{a}\text{T}_{b}\right)  }=\frac{r_{2n}^{\left(
\text{T}_{a}\right)  }+K^{\left(  \text{T}_{a}\text{T}_{b}\right)  }%
r_{2n}^{\left(  \text{T}_{b}\right)  }}{1+K^{\left(  \text{T}_{a}\text{T}%
_{b}\right)  }}, \label{r2nTaTb}%
\end{equation}
where
\begin{equation}
r_{2n}^{\left(  \text{T}\right)  }=c_{2n}^{\left(  \text{T}\right)  }p^{2n}.
\label{r2p}%
\end{equation}
For the Gaussian (G), disk (D), and ring (R):
\begin{subequations}
\label{c2n}%
\begin{align}
c_{2}^{\left(  \text{G}\right)  }  &  =2,\;c_{4}^{\left(  \text{G}\right)
}=12,\;c_{6}^{\left(  \text{G}\right)  }=120;\\
c_{2}^{\left(  \text{D}\right)  }  &  =1/4,\;c_{4}^{\left(  \text{D}\right)
}=1/8,\;c_{6}^{\left(  \text{D}\right)  }=5/64;\\
c_{2}^{\left(  \text{R}\right)  }  &  =1/2,\;c_{4}^{\left(  \text{R}\right)
}=3/8,\;c_{6}^{\left(  \text{R}\right)  }=5/16.
\end{align}

Defining derivatives $\mathfrak{r}_{2n}^{\left(  \text{R}\right)  }$ for the
annulus in Eq. (\ref{approxJ0}) using Eq. (\ref{r2n}), to leading order in the
small quantity $\varepsilon/p$,%
\end{subequations}
\begin{subequations}
\begin{align}
\mathfrak{r}_{2}^{\left(  \text{R}\right)  }/r_{2}^{\left(  \text{R}\right)
}  &  =1+\left(  1/4\right)  \left(  \varepsilon^{2}/p^{2}\right)  ,\\
\mathfrak{r}_{4}^{\left(  \text{R}\right)  }/r_{4}^{\left(  \text{R}\right)
}  &  =1+\left(  5/6\right)  \left(  \varepsilon^{2}/p^{2}\right)  ,\\
\mathfrak{r}_{6}^{\left(  \text{R}\right)  }/r_{6}^{\left(  \text{R}\right)
}  &  =1+\left(  7/4\right)  \left(  \varepsilon^{2}/p^{2}\right)  .
\end{align}
For $\varepsilon/p=0.1$, for the various quantities calculated in the text the
differences between the experimentally realizable annulus and the theoretical
ring are therefore less than a few percent, whereas for $\varepsilon/p=0.01$
these differences are of order $10^{-4}$.

As a specific example for, say, $r_{4}$, for a composite source with $a$ a
ring and $b$ a disk,
\end{subequations}
\begin{center}
$r_{4}^{\left(  \text{RD}\right)  }=\left(  3a^{4}/8+K^{\left(  \text{RD}%
\right)  }b^{4}/8\right)  /\left(  1+K^{\left(  \text{RD}\right)  }\right)  .$
\end{center}

\medskip

\section{Moments}

The moments $M_{2n}$ of the source function $S$ are defined for a composite
source by%
\begin{equation}
M_{2n}^{\left(  \text{T}_{a}\text{T}_{b}\right)  }=\dfrac{\int_{0}^{\infty
}u^{2n+1}S^{\left(  \text{T}_{a}\text{T}_{b}\right)  }\left(  u\right)
du}{\int_{0}^{\infty}uS^{\left(  \text{T}_{a}\text{T}_{b}\right)  }\left(
u\right)  du}.
\end{equation}
Expanding both $W\left(  pr\right)  $ and $J_{0}\left(  ur\right)  $ in Eq.
(\ref{VCZ}),%
\begin{equation}
M_{2n}^{\left(  \text{T}_{a}\text{T}_{b}\right)  }=\frac{\left(  n!\right)
^{2}4^{n}}{\left(  2n\right)  !}r_{2n}^{\left(  \text{T}_{a}\text{T}%
_{b}\right)  }. \label{M2W2}%
\end{equation}
It is useful to express $M_{2n}^{\left(  \text{T}_{a}\text{T}_{b}\right)  }$
for speckled speckle in a form that parallels Eq. (\ref{Wab}),
\begin{equation}
M_{2n}^{\left(  \text{T}_{a}\text{T}_{b}\right)  }=\frac{M_{2n}^{\left(
\text{T}_{a}\right)  }\left(  a\right)  +K^{\left(  \text{T}_{a}\text{T}%
_{b}\right)  }M_{2n}^{\left(  \text{T}_{b}\right)  }\left(  b\right)
}{1+K^{\left(  \text{T}_{a}\text{T}_{b}\right)  }}. \label{M2nTaTb}%
\end{equation}
where%
\begin{equation}
M_{2n}^{\left(  \text{T}\right)  }\left(  p\right)  =m_{2n}^{(\text{T})}%
p^{2n}, \label{Mmp}%
\end{equation}
with
\begin{equation}
m_{2}^{\left(  \text{T}\right)  }=2c_{2}^{\left(  \text{T}\right)  }%
,\;m_{4}^{\left(  \text{T}\right)  }=(8/3)c_{4}^{\left(  \text{T}\right)
},\;m_{6}^{\left(  \text{T}\right)  }=(16/5)c_{6}^{\left(  \text{T}\right)  }.
\label{mncn}%
\end{equation}
$c_{2n}^{\left(  \text{T}\right)  }$ is given in Eqs. (\ref{c2n}) for a
Gaussian (T = G), disk (T = D), and ring (T = R).

As a specific example, for a composite source with $a$ a Gaussian and $b$ a disk,

\begin{center}
$M_{6}^{\left(  \text{GD}\right)  }=\left(  384a^{6}+K^{\left(  \text{GD}%
\right)  }b^{6}/4\right)  /\left(  1+K^{\left(  \text{GD}\right)  }\right)  $.
\end{center}

Like the autocorrelation functions, the moments of composite sources are the
optical power weighted averages of the moments of the individual sources $a$
and $b$.

\medskip

\section{$\mathbf{\eta}$}

We discuss here the dimensionless parameter $\eta$ that plays a central role
in the level crossing statistics of derivatives of the phase, of the real and
imaginary parts of the wavefunction, and of the intensity. \ Extending [$17$],
we write for speckled speckle%
\begin{equation}
\eta^{\left(  \text{T}_{a}\text{T}_{b}\right)  }=r_{4}^{\left(  \text{T}%
_{a}\text{T}_{b}\right)  }/\left(  r_{2}^{\left(  \text{T}_{a}\text{T}%
_{b}\right)  }\right)  ^{2}-1, \label{eta}%
\end{equation}
where the normalized derivatives $r_{2n}^{\left(  \text{T}_{a}\text{T}%
_{b}\right)  }$ are given in Appendix B. \ Of special interest is the regime
where $\eta^{\left(  \text{T}_{a}\text{T}_{b}\right)  }$ becomes anomalously
large. \ We find that $\eta^{\left(  \text{T}_{a}\text{T}_{b}\right)  }$
reaches a maximum, $\eta_{\max}^{\left(  \text{T}_{a}\text{T}_{b}\right)  }$,
when $K^{\left(  \text{T}_{a}\text{T}_{b}\right)  }=\mathcal{K}_{\max
}^{\left(  \text{T}_{a}\text{T}_{b}\right)  }$, where
\begin{subequations}
\label{Etamax}%
\begin{align}
\mathcal{K}_{\max}^{\left(  \text{T}_{a}\text{T}_{b}\right)  }  &
=\frac{c_{2}^{\left(  \text{T}_{a}\right)  }}{c_{2}^{\left(  \text{T}%
_{b}\right)  }}\rho^{2},\label{etaKmax}\\
\eta_{\max}^{\left(  \text{T}_{a}\text{T}_{b}\right)  }  &  =\frac
{c_{4}^{\left(  \text{T}_{b}\right)  }}{4c_{2}^{\left(  \text{T}_{a}\right)
}c_{2}^{\left(  \text{T}_{b}\right)  }}\left(  \frac{1}{\rho^{2}}\right)  .
\label{Eta_max}%
\end{align}
For example, for $a$ a ring (R) and $b$\ a disk (D)%
\end{subequations}
\begin{align*}
\mathcal{K}_{\max}^{\left(  \text{RG}\right)  }  &  =\frac{c_{2}^{\left(
\text{R}\right)  }}{c_{2}^{\left(  \text{D}\right)  }}\rho^{2}=2\rho^{2},\\
\eta_{\max}^{\left(  \text{RG}\right)  }  &  =\frac{c_{4}^{\left(
\text{D}\right)  }}{4c_{2}^{\left(  \text{R}\right)  }c_{2}^{\left(
\text{D}\right)  }}\left(  \frac{1}{\rho^{2}}\right)  =\frac{1}{4\rho^{2}}.
\end{align*}

For normal speckle, for a Gaussian $\eta=2$, for a disk $\eta=1$, and for a
ring $\eta=\tfrac{1}{2}$. \ In contrast, for speckled speckle for say
$\rho=0.1$, $\eta_{\max}^{\left(  \text{GG}\right)  }=75$, $\eta_{\max
}^{\left(  \text{DD}\right)  }=50$, and $\eta_{\max}^{\left(  \text{RR}%
\right)  }=25$.

\begin{center}

{\large \textbf{References}}
\end{center}

\hspace{-0.15in}[1] J. W. Goodman, \emph{Speckle Phenomena In Optics} (Roberts
\& Co., Englewood, Colorado, 2007). \ A comprehensive review of the properties
of normal speckle is given here, together with a very extensive bibliography.

\hspace{-0.15in}[2] I. Freund and D. A. Kessler, \textquotedblleft
Singularities in speckled speckle,\textquotedblright\ Opt. Lett. \textbf{33,}
479-481 (2008).

\hspace{-0.15in}[3] D. A. Kessler and I. Freund, \textquotedblleft Short- and
long-range screening of optical phase singularities and C
points,\textquotedblright\ Opt. Commun. (in press).

\hspace{-0.15in}[4] M. Berry, \textquotedblleft Disruption of wave-fronts:
statistics of dislocations in incoherent Gaussian random
waves,\textquotedblright\ J. Phys. A \textbf{11,} 27-37 (1978).

\hspace{-0.15in}[5] B. I. Halperin, \textquotedblleft Statistical mechanics of
topological defects,\textquotedblright\ in Physics of Defects, R. Balian, M.
Kleman, and J.-P. Poirier, Eds. (North-Holland, Amsterdam, 1981), pp. 814-857.

\hspace{-0.15in}[6] N. B. Baranova, B. Ya Zel'dovich, A. V. Mamaev, N.
Pilipetskii and V. V. Shkukov, \textquotedblleft Dislocations of the
wave-front of a speckle-inhomogeneous field (theory and
experiment),\textquotedblright\ JETP Lett. \textbf{33,} 195-199 (1981).

\hspace{-0.15in}[7] M. R. Dennis, \textquotedblleft Phase critical point
densities in planar isotropic random waves,\textquotedblright\ J. Phys. A:
Math. Gen. \textbf{34,} L297-L303 (2003).

\hspace{-0.15in}[8] J. F. Nye and M. V. Berry \textquotedblleft Dislocations
in wave trains\textquotedblright, Proc. Roy. Soc. Lond. A \textbf{336,}
165-190 (1974).

\hspace{-0.15in}[9] J. F. Nye, J. V. Hajnal, and J. H. Hannay,
\textquotedblleft Phase saddles and dislocations in two-dimensional waves such
as the tides,\textquotedblright\ Proc. Roy. Soc. Lond. A \textbf{417,} 7-20 (1988).

\hspace{-0.15in}[10] I. Freund, \textquotedblleft Saddles, singularities, and
extrema in random phase fields,\textquotedblright\ Phys. Rev. E \textbf{52,}
2348-2360 (1995).

\hspace{-0.15in}[11] M. Born and E. W. Wolf, \emph{Principles of Optics}
(Pergamon Press, Oxford, 1959), Sect. 1.4.2.

\hspace{-0.15in}[12] J. F. Nye, \emph{Natural Focusing and Fine Structure of
Light} (IOP Publ., Bristol, 1999).

\hspace{-0.15in}[13] I. Freund, M. S. Soskin, and A. I. Mokhun,
\textquotedblleft Elliptic critical points in paraxial optical
fields,\textquotedblright\ Opt. Commun. \textbf{208}, 223-253 (2002).

\hspace{-0.15in}[14] I. Freund, \textquotedblleft Poincar\'{e}
vortices,\textquotedblright\ Opt. Lett. \textbf{26,} 1996-1998 (2001).

\hspace{-0.15in}[15] M. R. Dennis, \textquotedblleft Polarization
singularities in paraxial vector fields: morphology and
statistics,\textquotedblright\ Opt. Commun. \textbf{145,} 201--221 (2002);
\textquotedblleft Nodal densities of planar gaussian random
waves,\textquotedblright\ Eur. Phys. J. Special Topics \textbf{213,} 191-210 (2007).

\hspace{-0.15in}[16] I. Freund, \textquotedblleft Vortex
derivatives,\textquotedblright\ Opt. Commun. \textbf{137,} 118-126 (1997).

\hspace{-0.15in}[17] D. A. Kessler and I. Freund, \textquotedblleft
Level-crossing densities in random wave fields,\textquotedblright\ J. Opt.
Soc. Am. A \textbf{15,} 1608-1618 (1998).

\hspace{-0.15in}[18] I. Freund, \textquotedblleft`1001' correlations in random
wave fields,\textquotedblright\ Waves in Random Media \textbf{8,} 119-158 (1998).

\hspace{-0.15in}[19] D. E. Cartwright and M. S. Longuet-Higgins,
\textquotedblleft The statistical distribution of the maxima of a random
function,\textquotedblright\ Proc. R. Soc. Lond., Ser. A \textbf{237,}
212--232 (1956); M. S. Longuet-Higgins, \textquotedblleft Reflection and
refraction at a random moving surface: II. Number of specular points in a
Gaussian surface,\textquotedblright\ J. Opt. Soc. Am. \textbf{50,} 845--850 (1960).

\hspace{-0.15in}[20] M. V. Berry and J. H. Hannay, \textquotedblleft Umbilic
points on Gaussian random surfaces\textquotedblright, J. Phys. A \textbf{10,}
1809-1821 (1977).

\hspace{-0.15in}[21] S. O. Rice, \textquotedblleft Mathematical analysis of
random noise,\textquotedblright\ in \emph{Selected Papers on Noise and
Stochastic Processes}, N. Wax, Ed. (Dover, New York, 1954), pp. 133--294.
\ Although the planar optical field is two-dimensional, the density of zero
crossings along a straight line is a one-dimensional problem.

\end{document}